\documentclass[a4paper,11pt]{article}
\usepackage{jcappub}
\usepackage{amssymb}
\usepackage{graphicx}
\usepackage{subfigure}
\usepackage{epsfig,amsmath}
\usepackage{array}
\usepackage{multirow}
\usepackage{color}

\begin{document}

\title{Constraints on Modified Gravity models from White Dwarfs}

\author[a,1]{Srimanta Banerjee,\note{Corresponding author.}}
\author[b]{Swapnil Shankar}
\author[c]{and Tejinder P. Singh}

\affiliation[a,c]{Department of Astronomy and Astrophysics, Tata Institute of Fundamental Research, Mumbai 400005, Maharashtra, India}
\affiliation[b]{Department of Physics, Centre for Excellence in Basic Sciences, Mumbai 400098, Maharashtra, India}

\emailAdd{srimanta.banerjee@tifr.res.in}
\emailAdd{swapnil.shankar@cbs.ac.in}
\emailAdd{tpsingh@tifr.res.in}

\date{\today}

\abstract{Modified gravity theories can introduce modifications to the Poisson equation in the Newtonian limit. As a result, we expect to see interesting features of these modifications inside stellar objects. White dwarf stars are one of the most well studied stars in stellar astrophysics. We explore the effect of modified gravity theories inside white dwarfs. We derive the modified stellar structure equations and solve them to study the mass-radius relationships for various modified gravity theories. We also constrain the parameter space of these  theories from observations.}   
\keywords{modified gravity, white dwarfs}
\maketitle
\section{Introduction}
General relativity [GR] is an extremely successful theory and it has been verified by a variety of experimental tests. Nonetheless, there are fundamental issues, both theoretical and experimental, which suggest that possible classical generalisations of GR are worth investigating. Theoretical issues include singularity avoidance, both in gravitational collapse, and in the very early universe. The generic occurrence of such singularities is suggested by the singularity theorems, and it is important to investigate if modifications of GR, consistent with experiments, can overcome these singularities. On the observational side, the origin of cosmic acceleration and the flattening of galaxy rotation curves also pose a challenge to GR. While the standard cosmological model, i.e. $\Lambda CDM$, strongly favours a cosmological constant to explain cosmic acceleration, and cold dark matter particles to explain rotation curves, these explanations are not without their shortcomings. We do not properly understand why the observed value of the cosmological constant should be so much smaller than its theoretically favoured value, and yet be non-zero. Nor do we understand how vacuum energy couples to gravity. It could well be that explaining acceleration requires us to modify the law of gravitation on cosmological scales. As for cold dark matter, while there is excellent indirect evidence for it from structure formation, direct laboratory searches have not yielded any results so far. The case for modified gravity as an alternative to dark matter, though not a strong one yet, cannot be entirely discarded either. Modified theories of gravity also serve as important test-beds to analyse how well GR agrees with experiments.

Motivated by these reasons, a very large number of modified gravity theories have been proposed and their observational implications along with theoretical structures have been studied (see e.g. the review \cite{Clif} for a detailed discussion). Theories which succeed in avoiding singularities, or which serve as alternatives to dark energy / dark matter, must then be subjected to solar system tests, and tests in compact objects [binary pulsars, gravitational wave emission, periastron advance, white dwarf and neutron star physics etc.]. 

The present paper is concerned with testing of four modified gravity theories against the physics of white dwarfs. These theories were initially proposed to address particular issue(s) and later on discussed in other astrophysical or cosmological scenarios. The scalar-vector-tensor gravity theory (STVG), also referred to as Modified Gravity (MOG) in the literature, was proposed by Moffat \cite{mof1} as an extension to non-symmetric gravity theory (NGT) and the metric skew-tensor gravity theory (MTG), to explain flattening of galaxy rotation curves without invoking dark matter \cite{mof1}, \cite{mof2}. Later on Moffat and his collaborators studied the theory in the context of cluster dynamics \cite{mof3}, Bullet Cluster \cite{mof4} and cosmology \cite{mof1}, \cite{mof6} without considering the contributions from the dark matter component. Recently this theory has also been considered in the context of neutron stars \cite{mof5} and recent observation of gravitational waves \cite{mof7}. In all these settings the observations are claimed to be in good agreement with the theory. 

As we mentioned in the beginning, GR is plagued by the formation of singularities which signal a breakdown of the theory. Thus higher curvature corrections may be important to 
address this particular issue. Eddington-inspired Born-Infeld gravity, proposed by Ba\~nados and Ferraira \cite{Ban}, was shown to remove singularity in the early universe \cite{Ban} as well as the gravitational collapse of dust particles \cite{Pani}. This theory was initially proposed as a gravitational analogue of Born-Infeld electrodynamics and it was found to contain higher order matter couplings besides being identical to GR outside matter \cite{Ban}, \cite{Pani}. The astrophysical aspects of this theory have been studied in \cite{Cardoso}, \cite{Avelino} and the cosmological consequences are analysed in \cite{Scar},  \cite{Avelino}. But this theory is plagued by surface singularities \cite{Soti} which puts it on a somewhat shaky ground, although  gravitational backreaction \cite{Kim} can be shown to rescue the theory from this pathological situation.

The history of considering higher order derivative terms in the action (hence in field equations) is quite old and it emerged from attempts by Weyl and Eddington to include electromagnetic fields to obtain a unified framework. Later on it was shown that higher derivative terms can improve  renormalizability properties \cite{Stella2}, but the theory could become vulnerable to ghosts or instabilities \cite{Clif}. Several realistic fourth order gravity theories \cite{Clif}, \cite{Tsuji} have been proposed and studied in great detail in the context of inflation, dark energy or dark matter. In this study we would particularly consider two fourth order modified gravity models, one being a particular type of $f(R)$ gravity theory \cite{Tsuji} and the other a particular case of quadratic gravity theory \cite{Stella} (we will be referring to the latter as fourth order gravity  (FOG) in the present study although $f(R)$ gravity is also a fourth order gravity theory). We would also consider two other fourth order gravity models which share the same Newtonian limit as the FOG model mentioned above despite having different field equations. One of these  fourth order gravity models was proposed to explain flattening of rotation curves \cite{Priti} without dark matter and late time cosmic acceleration \cite{Shreya}. Another one was proposed as an effective theory of gravity where the correction terms in Einstein field equations originate due to the consideration of the effect of induced gravitational polarization \cite{Zal}. 

All these four theories STVG, EiBI, FOG or $f(R)$ give rise to novel physics in weak field regime because of the presence of new terms. White dwarf stars are known to be well understood in the weak field regime of GR as the strong field effects ($\frac{GM}{c^2 \mathcal{R}}\sim 10^{-4}$) are small in these stellar objects. The modified gravity theories introduce new repulsive or attractive terms in the weak field Newtonian limit and all these terms affect the physics of white dwarf stars through the stellar structure equations. As a result, the mass-radius relation for these stars gets modified. We will study the effects of these additional terms and constrain their effects from observations. The discussion of white dwarf stars in the context of modified gravity theory was pursued earlier in \cite{Jain}, \cite{Banim} although they have considered theories different from ours.

The plan of this paper is as follows. In Sec. \ref{m}, we briefly review the basic features of the four modified gravity theories mentioned above and we also obtain the Newtonian limit of these theories. In Sec. \ref{n}, we give a brief exposition to the physics of white dwarf stars and recall the physics relevant to this work. Sec. \ref{o} is completely devoted to the results. Here we first discuss the formalism and the numerical scheme that has been employed in this work. Then we explore the phenomenology of these theories by considering the mass radius relation for white dwarfs and  impose constraints on the parameter space of these theories from observations. We draw our conclusions and discuss the future prospects in Sec \ref{p}.

\section{Modified gravity theories}\label{m}
We discuss the basic features of the modified gravity theories that we are considering for this work. Then we obtain Newtonian limit of these modified gravity theories as we need the expression for radial acceleration inside a spherically symmetric object for solving the stellar structure equations of white dwarfs. We use $\hbar=c=1$ and the metric signature $\left(-1,+1,+1,+1\right)$ throughout the paper.  

Before going into the details of modified gravity theories let us briefly recall the crux of Newtonian limit of any relativistic gravitation theory \cite{Strau}. In static weak field and slow motion limit or in Newtonian limit of any gravitation theory we basically consider all the terms in the field equation up to the order of 
$v^2\ (v^2\ll1)$. We use Einstein gravity as a prototype theory to recall the effect of this approximation
\begin{equation}
R_{ab}-\frac{1}{2}g_{ab}R=8\pi G T^{(m)}_{ab}
\label{a}
\end{equation}
where $R_{ab}$, $R$ and $T^{(m)}_{ab}$ are Ricci tensor, Ricci scalar and energy momentum tensor respectively. In order to study Newtonian limit let us perturb the metric about the Minkowski space
\begin{equation}
g_{ab}\simeq\eta_{ab}+g^{(2)}_{ab}
\end{equation}
where $\eta_{ab}$ is the Minkowski metric and $g^{(2)}_{ab}$ is the first order correction to Minkowski metric of the order $v^2$. The inverse of $g_{ab}$ also can be shown to be given by
\begin{equation}
g^{ab}\simeq\eta^{ab}-g^{(2){ab}}
\end{equation}
Now if one computes the $00$ component of Ricci tensor in this limit, one essentially arrives at \cite{Strau}
\begin{equation}
R_{00}^{(2)}=-\frac{1}{2}\nabla^2 g^{(2)}_{00}
\label{b}
\end{equation}
assuming $g^{(2)}_{00}$ to be static. Also the energy momentum tensor $T^{(m)}_{ab}$ assumes the form of dust $=\rho u_a u_b$ as the effect of pressure may be neglected in the Newtonian limit. Therefore $T=g^{ab}T^{(m)}_{ab}=-\rho$. Hence putting the above result (\ref{b}) into $00$ component of (\ref{a}) and considering the trace equation of Einstein field equation $(R=-8\pi G T)$, one essentially obtains \cite{Strau}
\begin{equation}
\nabla^2 g^{(2)}_{00}=-8\pi G \rho
\end{equation} 
Hence comparing the above with Poisson equation, one can identify $g^{(2)}_{00}=-2\Phi$ where $\Phi$ is the Newtonian potential.  
In case of modified gravity theories we obtain an effective Newtonian potential $\Phi$ following the above method due to additional terms in their field equations. Hence  Poisson equation would be modified, obtaining contributions from the additional terms in Newtonian limit. We explore and constrain the observational imprints of these additional terms by studying  white dwarf stars.  
\subsection{Scalar-Tensor-Vector gravity}\label{stvg}
The generic form of the action for the Scalar-Tensor-Vector gravity (STVG) is given by \cite{mof1,mof2,mof3}
\begin{equation}
S=S_G+S_{\phi}+S_s+S_m
\end{equation}
where $S_G$ is the usual Einstein-Hilbert action with cosmological constant $\Lambda$,
\begin{equation}
S_{G}=\frac{1}{16\pi}\int\frac{1}{G}\left(R-2\Lambda\right)\sqrt{-g}d^4x
\label{11a}
\end{equation}
$S_{\phi}$ is the action for the massive vector field $\phi^{a}$
\begin{equation}
S_{\phi}=-\frac{1}{4\pi}\int\omega\left(\frac{1}{4}B^{ab}B_{ab}-\frac{1}{2}\mu^2\phi^a\phi_a+V_{\phi}\left(\phi_{a} \phi^{a}\right)\right)  \sqrt{-g}d^4x
\end{equation}
$S_s$ is the action for the scalar fields $G$ and $\mu$.
\begin{equation}
S_{s}=-\int\frac{1}{G}\left[\frac{1}{2}g^{ab}\left(\frac{\nabla_a G \nabla_b G}{G^2}+\frac{\nabla_a \mu \nabla_b \mu}{\mu^2}\right)
+\frac{V_G(G)}{G^2}+\frac{V_\mu(\mu)}{\mu^2}\right]\sqrt{-g}d^4x
\end{equation}
and $S_m$ is the action for the matter field.
Here, $B_{ab}=\partial_a\phi_b-\partial_b\phi_a$ is a skew symmetric tensor field with $\phi_a$ playing the role of the vector field and the gravitational constant $G$ is a scalar field. The scalar field $\mu$ represents the mass of the vector field $\phi^{a}$ and $\omega$ is the dimensionless coupling constant. $V_\phi$, $V_G$ and $V_\mu$ are the self-interaction potentials for the vector and scalar fields. In this work we would ignore the contribution of the potentials and cosmological constant. The action for the massive vector field resembles the form of Maxwell-Proca field whereas the action for the gravitational constant has got the Brans-Dicke form.

The variation of the STVG action w.r.t. $g^{ab}$ gives \cite{mof1}
\begin{equation}
G_{ab}+Q_{ab}=8\pi G T_{ab}
\label{1a}
\end{equation}
where, 
$Q_{ab}=G\left(\square \frac{g_{ab}}{G}-\nabla_a \nabla_b \frac{1}{G} \right)$ and $T_{ab}=T_{ab}^{(m)}+T_{ab}^{(\phi)}+T_{ab}^{(s)}$
is the total energy momentum tensor. Also $\square=g^{ab}\nabla_a\nabla_b$ is the D'Alembertian operator. $T_{ab}^{(m)}$, $T_{ab}^{(\phi)}$ and $T_{\alpha\beta}^{(s)}$ are the energy momentum tensors for ordinary matter field, vector field and scalar fields respectively
\begin{equation}
\frac{-2}{\sqrt{-g}}\frac{\delta S_m}{\delta g^{ab}}=T_{ab}^{(m)}, 
 \frac{-2}{\sqrt{-g}}\frac{\delta S_\phi}{\delta g^{ab}}=T_{ab}^{(\phi)},
 \frac{-2}{\sqrt{-g}}\frac{\delta S_s}{\delta g^{ab}}=T_{ab}^{(s)}
\end{equation}
The expressions for the different energy momentum tensors are given in \cite{mof3}. 

Apart from metric tensor, matter is also coupled to the vector field which gives rise to fifth force and this is captured by matter current density $J^a$
\begin{equation} 
J^a=-\frac{1}{\sqrt{-g}}\frac{\delta S_m}{\delta\phi_a}
\end{equation}
The variation of the action w.r.t. $\phi_b$ gives \cite{mof1},
\begin{equation}
\nabla_a B^{ab}-\mu^2\phi^b=-\frac{4\pi J^b}{\omega}
\label{1b}
\end{equation}
There are also field equations for the scalar fields which are given in \cite{mof1,mof7}.

As the matter is also coupled to the massive vector field, test particles would not follow geodesics; there will be a fifth force term in the equation. The equation of motion for a test particle in STVG can be shown to be given by \cite{mof1,mof2}
\begin{equation}
m\left(\frac{d^2x^c}{d\tau^2}+\Gamma^{c}_{ab}\frac{dx^a}{d\tau}\frac{dx^b}{d\tau}\right)=f^c
\label{1e}
\end{equation} 
where $f^c=\lambda\omega B^{a}_{\ c} \frac{dx^c}{d\tau}$. Here $\tau$ is the affine parameter along the trajectory of the particle and $\lambda$ which is the fifth force charge of the test particle is defined by $\lambda=\kappa m$, $m$ being the mass of the test particle and $\kappa$ being the coupling constant. Since the fifth force charge depends on the test particle mass, the equation for test particle becomes mass independent. Hence in STVG, although there is fifth force, weak equivalence principle is not violated.

Linearising the vector field equation (\ref{1b}) about the Minkowski space, one arrives at \cite{mof2}
\begin{equation}\label{vec}
\nabla^2\phi_0-\mu^2\phi_0=-\frac{4\pi J^0}{\omega}
\end{equation}
in the weak field static limit. Here, we have assumed the conservation of $J^a$ along with the gauge condition of $\phi^a$. Similarly the spatial divergence of geodesic equation (\ref{1e}) in this limit gives \cite{mof2}
\begin{equation}
\overrightarrow{\nabla}.\overrightarrow{a}-\frac{1}{2}\nabla^2h_{00}=-\omega\kappa\nabla^2\phi_0
\label{geo}
\end{equation}
following the method discussed in the beginning of this section. We have used $J^0=\kappa\omega\rho$ in arriving at the last expression. The $00$ component of (\ref{1a}) can be shown to produce exactly the equation (\ref{b}) in this limit upon using the approximations that the density of vector fields is small compared to that of matter fields, $\mu$ is constant and $G$ would assume the background value \cite{mof2} along with ignoring  higher order perturbations in vector field. Hence one can obtain the modified Poisson equation by combining the Newtonian limit of field equations (\ref{vec}) and geodesic equation (\ref{geo})
\begin{equation}\label{poistvg}
\nabla^2\Phi=4\pi G \rho+\kappa\omega\nabla^2\phi_0
\end{equation}
By solving the above one arrives at the expression for effective potential $\Phi\left(\textbf{r}\right)$ \cite{mof2}
\begin{equation}
\Phi\left(\textbf{r}\right)=-G_N\left(1+\alpha\right)\int\frac{\rho\left(\textbf{r}'\right)}{\mid\textbf{r}-\textbf{r}'\mid}d^3r'+G_N\alpha\int\frac{\rho\left(\textbf{r}'\right)}{\mid\textbf{r}-\textbf{r}'\mid}e^{-\mu\mid\textbf{r}-\textbf{r}'\mid}d^3r'
\label{1f}
\end{equation}
where $\alpha=\frac{G_{\infty}-G_N}{G_N}=\kappa^2 G_N^{-1}$, $G_N$ and $G_\infty$ being the Newtonian gravitational constant, and effective gravitational constant at infinity, respectively. 
  
Let us now closely analyse the equation for the effective potential that we have got in Newtonian limit. The effective potential is endowed with an attractive Newtonian term which gets enhanced by a factor $(1+\alpha)$ besides having a repulsive Yukawa term (with a factor $\alpha$) which emerged from the massive vector field. The usual attractive Newtonian as well as the Yukawa term got the enhanced factor from the scalar field $G$. The interplay between the enhanced attractive part and repulsive Yukawa part captures the essence of the theory and by suitably tweaking the parameters one can successfully explain various astrophysical observations. 

The effective potential and radial acceleration inside a spherically symmetric object of radius $\mathcal{R}$ are given by
\begin{eqnarray}
&\Phi\left(r\right)&=-\frac{4\pi G_N\left(1+\alpha\right)}{r}\int_{0}^{r}r'^2\rho(r')dr'-4\pi G_N\left(1+\alpha\right)\int_r^{\mathcal{R}}r'\rho(r')dr'\nonumber\\&+&                           \frac{4\pi G_N\alpha}{\mu r}e^{-\mu r}\int_{0}^{r}r'\rho(r')\sinh\left(\mu r'\right)dr'+\frac{4\pi G_N\alpha}{\mu r}\sinh\left(\mu r\right)\int_{r}^{\mathcal{R}}r'\rho(r')e^{-\mu r'}dr'
\end{eqnarray}
and
\begin{eqnarray}
a\left(r\right)&=&-\frac{d\Phi}{dr}\nonumber\\
&=&-\frac{4\pi G_N\left(1+\alpha\right)}{r^2}\int_{0}^{r}r'^2\rho(r')dr'+\frac{4\pi G_N\alpha}{\mu r^2}\left(1+\mu r\right)e^{-\mu r}\int_{0}^{r}r'\rho(r')\sinh\left(\mu r'\right)dr'\nonumber \\&&+\frac{4\pi G_N\alpha}{\mu r^2}\left[\sinh\left(\mu r\right)-\mu r \cosh\left(\mu r\right)\right]\int_{r}^{\mathcal{R}}r'\rho(r')e^{-\mu r'}dr'
\label{STVG}
\end{eqnarray}
So we see that Gauss' law is violated here and this violation is essentially due to the presence of Yukawa term in the potential.
\subsection{Eddington inspired Born-Infeld gravity}\label{eibi}
Eddington inspired Born-Infeld gravity (EiBI) is described by the action \cite{Ban}
\begin{equation}
S=\frac{1}{8\pi G_N \kappa}\int d^4x\left(\sqrt{\mid g_{ab}+\kappa R_{ab}(\Gamma)\mid}-\lambda\sqrt{-g}\right)+S_m\left(g_{ab},\chi_m\right)
\label{2a}
\end{equation}
where $\mid.\mid$ represents determinant, $\kappa$ is the independent parameter in the theory, $\lambda(\neq0)$ is a dimensionless constant, $R_{ab}$ is the symmetric part of Ricci tensor built from the connection $\Gamma^c_{ab}$ and $S_m\left(g_{ab},\chi_m\right)$ is the matter action with $\chi_m$ representing any matter field. The above action produces Einstein-Hilbert action (\ref{11a}) with cosmological constant $\Lambda=\frac{\left(\lambda-1\right)}{\kappa}$ in the small curvature limit $\kappa R\ll1$ whereas it tends to Eddington action in the limit $\kappa R\gg1$ which is given by \cite{Ban,Pani}
\begin{equation}
S_{Edd}=\frac{\kappa}{8\pi G_N}\int \sqrt{\mid R_{ab}\mid} d^4x
\end{equation} 
Hence one expects to see novel features of the theory in the high density region like neutron stars or early universe.

In EiBI gravity the metric and connection are considered as independent fields as in Palatini's approach in Einstein gravity, giving a hint to the bimetric structure of the theory. The metric approach is shown to be plagued by ghosts which can be eliminated by adding higher order terms in the action \cite{Desar,Pani}. Also matter is minimally coupled to metric tensor only in this theory.

The variation of the EiBI action (\ref{2a}) w.r.t. $\Gamma^c_{ab}$ gives \cite{Ban}
\begin{eqnarray}
q_{ab}&=&g_{ab}+\kappa R_{ab}(q) \nonumber \\
\Gamma^c_{ab}&=&\frac{1}{2}q^{cd}\left(\partial_a q_{bd}+\partial_b q_{ad}-\partial_d q_{ab}\right)
\label{2b}
\end{eqnarray}
where $q_{ab}$ is an auxiliary metric compatible with the connection $\Gamma$.

By varying the action (\ref{2a}) w.r.t. $g_{ab}$ one arrives at \cite{Ban}
\begin{equation}
\sqrt{-q}q^{ab}=\lambda\sqrt{-g}g^{ab}-8\pi G_N\kappa\sqrt{-g}T^{(m)ab}
\label{2c}
\end{equation}
where $q^{ab}$ is the inverse of $q_{ab}$ and $T^{(m)}_{ab}$ is the usual energy momentum tensor which satisfies the conservation law i.e. $\nabla^{a} T^{(m)}_{ab}=0$ because of the usual coupling between matter fields and metric tensor. The theory can be shown to be equivalent to Einstein gravity in the absence of matter. Hence all the observations done in free space would identically hold for EiBI gravity also. Also as the auxiliary metric is connected to the metric tensor algebraically, this theory has got only metric tensor as the dynamical field. 

Expanding the field equations (\ref{2b}), (\ref{2c}) in powers of $\kappa$, we get \cite{Ban}
\begin{equation}
R_{ab}\left(\Gamma\right)\simeq \Lambda g_{ab}+ 8\pi G_N\left( T_{ab}^{(m)}-\frac{1}{2}T^{(m)} g_{ab}\right)+8\pi G_N\kappa\left[S_{ab}-\frac{1}{4}Sg_{ab}\right]+\mathcal{O}\left(\kappa^2\right)
\label{2d}
\end{equation}
where $S_{ab}=T^{(m)c}_aT_{cb}^{(m)}-\frac{1}{2}T^{(m)}T_{ab}^{(m)}$. The above quadratic corrections to the matter fields have an uncanny resemblance to the induced field equations on the brane, in Shiromizu-Maeda-Sasaki approach \cite{Sasaki,Roy}. Thus EiBI theory is endowed with non-trivial corrections to GR inside matter whilst remaining identical to Einstein gravity outside matter. These lowest order corrections leave a rich imprint on the Newtonian limit giving the modified Poisson equation \cite{Ban,Pani}
\begin{equation}
\nabla^2 \Phi=4\pi G_N \rho+\frac{\kappa}{4}\nabla^2\rho
\end{equation}  
which we have got by linearising the equations (\ref{2c}), (\ref{2b}) about the Minkowski space following the method described in the beginning of Sec. (\ref{m}). Hence the expression for radial acceleration is given by
\begin{equation}
a\left(r\right)=-\frac{G_N m(r)}{r^2}-\frac{k}{4}\frac{d\rho}{dr}
\label{EiBI}
\end{equation}
Inside a stellar object the term $\frac{d\rho}{dr}$ is negative. Thus the correction term acts as repulsive force and it can be shown that it corresponds to an effective polytropic fluid with equation of state $P_{\text{eff}} = K\rho^2$ where $K=\kappa/8$ \cite{Pani}. 
\subsection{Fourth order gravity theories}\label{fog} 

In this section we discuss a modified gravity model emerging as a particular case of the quadratic gravity theory proposed by Stella \cite{Stella}. The Newtonian limit of this model will be shown to produce the biharmonic modification to the usual Poisson equation which is reminiscent of the Bopp-Podolsky theory in nonlinear electrodynamics \cite{Bop}, \cite{Pod}. This particular modified Poisson equation can be shown to emerge as a consequence of considering the effect of quadrupole gravitational polarization which serves the physical motivation for discussing only this particular case. This connection is quite intriguing and was considered first by \cite{Bel1}.  
We develop this relationship, motivated by the averaging problem in macroscopic gravity, and also discuss the crux of physical arguments considered in \cite{Bel1}.

The action for quadratic gravity is given by \cite{Stella}
\begin{equation}
S=\frac{1}{16\pi G_N}\int \sqrt{-g}d^4x\left(R+\gamma R^2+\beta R^{ab}R_{ab}\right)+S_m(g_{ab},\chi_m)
\end{equation}
where $\gamma$, $\beta$ are the dimensionless parameters and $S_m(g_{ab},\chi_m)$ is the action for any matter field $\chi_m$. The variation of the above action w.r.t. $g^{ab}$ is given by \cite{Stella}
\begin{eqnarray}
8\pi G_N T^{(m)}_{ab}&=&-2\gamma\nabla_a\nabla_b R+\beta\square R_{ab}+\left(\frac{\beta}{2}+2\gamma\right)g_{ab}\square R+2\beta R^{c}_{a}R_{bc}\nonumber\\&+&2\gamma R R_{ab}-\frac{g_{ab}}{2}\left(\beta R^{cd}R_{cd}+\gamma R^2\right)-2\beta\nabla_{bc}R^{c}_{a}+R_{ab}-\frac{1}{2}g_{ab}R
\label{fog1}
\end{eqnarray}  
The equation reduces to Einstein field equation in the limit $\gamma\rightarrow0$ and $\beta\rightarrow0$. Since matter is only coupled to the metric tensor minimally, the energy momentum tensor satisfies the conservation law  $\nabla^{a} T^{(m)}_{ab}=0$. The gravitational field in this theory has eight degrees of freedom, two representing the massless graviton, one the massive scalar and remaining five describing massive graviton modes \cite{Stella}.

Linearising the field equation (\ref{fog1}) about the Minkowski space following the method mentioned in the beginning of section (\ref{m}), we obtain \cite{Clif}
\begin{eqnarray}
2(3\gamma+\beta)\nabla^2 R-R&=-&8\pi G_N \rho\\
(4\gamma+\beta)\nabla^2R-R-2\nabla^2(\Phi+ \beta\nabla^2\Phi) &=&-16\pi G_N\rho
\end{eqnarray} 
For the case $2\gamma+\beta=0$, it takes the following form \cite{Schmidt}
\begin{equation}
\nabla^2\Phi-2\gamma\nabla^4\Phi=4\pi G_N \rho
\end{equation}
which we are going to use in this study. This case has also been studied in \cite{Havas}.
If one considers a medium composed of self-gravitating objects deformable by tidal effects arising due to the inhomogeneities of the global field, one can show that the gravitational potential inside that medium follows the above equation in the continuum limit. As a result of this tidal effect, quadrupolar polarisation originates in the medium which is otherwise composed of mass monopoles. The density of the quadrupoles can be shown to introduce the biharmonic term for small deformations of the medium  and the parameter $\gamma$ depends upon the quadrupolar deformability of the objects composing the medium. We refer  the reader to \cite{Bel1} for more details. 

The choice $3\gamma+\beta=0$ can be shown to produce Weyl squared modification to Einstein gravity \cite{Frolov} and this particular case has also been considered in non-commutative spectral gravity \cite{Mairi}.

The fourth order modification to Poisson equation can also be shown to emerge as a consequence of considering the effect of induced gravitational polarization in a macroscopic medium \cite{Zal}. The averaged field equation for any macroscopic object in continuum limit, starting from the Einstein field equation of that object in microscopic description where the object is considered as a collection of `molecules', is given by \cite{Zal}
\begin{equation}
R_{ab}-\frac{1}{2}g_{ab}R=8\pi G_N(T_{ab}^{(free)}+T_{ab}^{(GW)}+\frac{1}{2}\nabla^d\nabla^cQ_{acbd})
\end{equation}
where one has modeled the effect of gravitational polarization using Szekeres' approach \cite{Sze}. Here `molecules' represent the clump of microscopic particles bound by gravitational attraction and the quantity $Q_{acbd}$ is the gravitational quadrupole polarization tensor which according to Szekeres' model can be assumed to take the form $Q_{i0j0}=\epsilon_g R_{i0j0}$ where $\epsilon_g$ is the gravitational dielectric constant and $R_{abcd}$ is the usual Riemann tensor \cite{Zal}. The quantity $T_{ab}^{(GW)}$ is the Isaacson's energy momentum tensor \cite{Isaac} and it is connected to traceless part of gravitational quadrupole polarization tensor $Q_{abcd}$ \cite{Zal}. Also $T_{ab}^{(free)}$ is the energy momentum tensor of the `molecules'. All the energy momentum tensors along with the term containing polarization tensor can be shown to be divergence-less. The basic idea in this formalism is that the gravitational field inside a `molecule' (can be thought of as a galaxy) gets modified because of induced polarization due to other `molecules' in a macroscopic object (can be thought of as `galaxy cluster') following the same ideas as in electrodynamics. Now one considers Isaacson's averaging procedure in order to study the effect of averaging over Einstein field equations which produces the additional two terms appearing in the averaged macroscopic equation. In Newtonian limit above model can be shown to give \cite{Zal}
\begin{equation}
\nabla^4\Phi-\zeta^2\nabla^2\Phi=-4\pi G_N\zeta^2 \rho
\label{fog2}
\end{equation}
where $\zeta^2=\frac{3}{4\pi G_N \epsilon_g}$ and also in Newtonian limit $T_{ab}^{(GW)}$ does not contribute. Recently one modified gravity model \cite{Priti} was proposed taking the inspiration from the effect of induced gravitational polarization. The field equation of the model is given by \cite{Priti}
\begin{equation}
R^{ab}-\frac{1}{2}g^{ab}R=8\pi G_N T^{(m)ab}+\zeta^{-2}\nabla_c\nabla_d R^{acbd}
\end{equation} 
It was shown to explain the late time cosmic acceleration along with the rotational curves of galaxies without invoking dark matter \cite{Priti, Shreya}. The model also produces the same modified Poisson equation (\ref{fog2}) in Newtonian limit \cite{Priti}.

Thus we have found that both the modified gravity models discussed above although studied in different contexts give rise to same biharmonic form in Newtonian limit. Therefore the expression for the effective Newtonian potential can be obtained by solving the equation (\ref{fog2}); it gives (see \cite{Zal})
\begin{equation}
\Phi\left(\textbf{r}\right)=-G_N\int\frac{\rho\left(\textbf{r}'\right)}{\mid\textbf{r}-\textbf{r}'\mid}d^3r'+G_N\int\frac{\rho\left(\textbf{r}'\right)}{\mid\textbf{r}-\textbf{r}'\mid}e^{-\zeta\mid\textbf{r}-\textbf{r}'\mid}d^3r'
\label{fog3}
\end{equation}
So we again obtain the repulsive Yukawa term. Also the equation is nearly identical to the effective potential (\ref{1f}) for STVG excepting the fact that in STVG the strength of the attractive term as well as the repulsive term gets multiplied by a factor of $\alpha$ which comes from the consideration of gravitational constant as a scalar field. Hence the expressions for effective potential and radial acceleration for a spherically symmetric object of radius $\mathcal{R}$ are given by
\begin{eqnarray}
\Phi\left(r\right)&=&-\frac{4\pi G_N}{r}\int_{0}^{r}r'^2\rho(r')dr'-4\pi G_N\int_r^{\mathcal{R}}r'\rho(r')dr'\nonumber\\&+&\frac{4\pi G_N}{\zeta r}e^{-\zeta r}\int_{0}^{r}r'\rho(r')\sinh\left(\zeta r'\right)dr'+\frac{4\pi G_N}{\zeta r}\sinh\left(\zeta r\right)\int_{r}^{\mathcal{R}}r'\rho(r')e^{-\zeta r'}dr'
\end{eqnarray}
and
\begin{eqnarray}
a\left(r\right)&=&-\frac{d\Phi}{dr}\nonumber\\
&=&-\frac{4\pi G_N}{r^2}\int_{0}^{r}r'^2\rho(r')dr'+\frac{4\pi G_N}{\zeta r^2}\left(1+\zeta r\right)e^{-\zeta r}\int_{0}^{r}r'\rho(r')\sinh\left(\zeta r'\right)dr'\nonumber \\&&+\frac{4\pi G_N}{\zeta r^2}\left[\sinh\left(\zeta r\right)-\zeta r \cosh\left(\zeta r\right)\right]\int_{r}^{\mathcal{R}}r'\rho(r')e^{-\zeta r'}dr'
\label{FOG}
\end{eqnarray}
\subsection{$f(R)$ gravity}\label{fr}
The action for the $f(R)$ gravity is given by \cite{Tsuji}
\begin{equation}
S=\frac{1}{16\pi G_N}\int d^4x \sqrt{-g}f\left(R\right)+S_m(g_{ab},\chi_m)
\end{equation}
where $S_m$ is the action for any matter fields $\chi_m$ and $f(R)$ is an arbitrary function of the Ricci scalar $R$.

The field equation for the $f(R)$ gravity can be obtained by varying the action w.r.t. $g^{ab}$ \cite{Tsuji}
\begin{equation}
f'\left(R\right)R_{ab}-\frac{1}{2}f(R)g_{ab}-\nabla_a \nabla_b f'(R)+g_{ab}\square f'(R)=8\pi G_N T^{(m)}_{ab}
\label{3a}
\end{equation}
where $f'(R)=\frac{df(R)}{dR}$ and $T^{(m)}_{ab}$ is the energy momentum tensor for the matter fields. Since matter is only coupled to metric tensor minimally, $T^{(m)}_{ab}$ satisfies the conservation law i. e. $\nabla^{a} T^{(m)}_{ab}=0$. The trace of (\ref{3a}) gives \cite{Tsuji}
\begin{equation} 
3\square f'(R)+f'(R)R-2f(R)=8\pi G_N T^{(m)}
\label{3b}
\end{equation}
where $T=g^{ab}T^{(m)}_{ab}$. So we see that $f'(R)$ is a dynamical scalar field in $f(R)$ gravity due to presence of the term $\square f'(R)$ in equation (\ref{3b}). One can easily arrive at Einstein gravity by putting $f(R)=R$ into the equations (\ref{3a}) and (\ref{3b}) for which $\square f'(R)$  vanishes.

In this study instead of considering any particular $f(R)$ gravity model we focus on a generic form of $f(R)$ gravity model where the function $f(R)$ is analytically Taylor expandable about a certain value $R=R_0$ \cite{Clifton,Cappo} 
\begin{eqnarray}
f(R)&=& \sum \limits_n ‎\frac{f^n(R_0)‎}{n!} (R-R_0)^n \nonumber \\
&=& c_0+c_1 R+c_2 R^2+c_3 R^3+....
\label{3c}
\end{eqnarray}
where $f^n(R)$ represents the n-th derivative of $f(R)$ w.r.t. $R$ and $c_0$ essentially correspond to cosmological term which we set to zero as we are assuming the space-time to be asymptotically Minkowski (i.e. $R_0=0$). We also assume $c_1=1+\delta$ where $\delta$ is an independent parameter describing the deviation from Einstein gravity value of $c_1$ which may acquire non-trivial values on astronomical scales \cite{Cappo}. It also must take a value in the range $-1<\delta$ otherwise gravity would become repulsive. Also $c_2$ must be positive in order to avoid tachyonic instability \cite{Tsuji}.  

The Newtonian limit of these theories was first obtained by \cite{Sexl} for a point mass. Here we follow the derivations carried out in \cite{Clifton}. Linearising the trace equation (\ref{3b}) about the Minkowski space and using the form of $f(R)$ given in (\ref{3c}) we arrive at \cite{Clifton}
\begin{equation}
\left(\nabla^2-\xi^2\right)R^{(2)}=-\frac{8\pi G_N\xi^2}{1+\delta}\rho
\end{equation}
where $\xi=\sqrt{\frac{c_1}{6c_2}}$ defines the mass of the scalar field $R^{(2)}$, $R^{(2)}$ being the Ricci scalar up to $\mathcal{O}(2)$ and $\delta=c_1-1$. The solution of the above inhomogeneous Helmholtz equation gives \cite{Clifton}
\begin{equation}
R^{(2)}=\frac{2G_N\xi^2}{1+\delta}\int \frac{\rho(\textbf{r}')}{\mid \textbf{r}-\textbf{r}'\mid}e^{-\xi \mid \textbf{r}-\textbf{r}'\mid}d^3r'
\label{3d}
\end{equation}
Also from the field equation (\ref{3a}) and the trace equation (\ref{3b}) we get \cite{Clifton}
\begin{eqnarray}
\nabla^2\left(\frac{c_1}{4}g_{00}^{(2)}+\frac{c_1}{4}g_{ii}^{(2)}+2c_2 R^{(2)}\right)&=&-8\pi\rho G \\
\nabla^2\left(c_1g_{ii}^{(2)}+5c_1g_{00}^{(2)}\right)&=&-64\pi\rho G
\end{eqnarray}
Combining the above equations we arrive at the modified Poisson equation for $f(R)$ gravity
\begin{equation}
\nabla^2\Phi(\textbf{r})=\frac{4\pi G}{1+\delta}\rho(\textbf{r})-\frac{1}{6\xi^2}\nabla^2R^{(2)}
\end{equation}
Hence the effective potential $\Phi$ is given by
\begin{equation}
\Phi\left(\textbf{r}\right)=-\frac{G_N}{1+\delta}\int\frac{\rho\left(\textbf{r}'\right)}{\mid\textbf{r}-\textbf{r}'\mid}d^3r'-\frac{G_N}{3 (1+\delta)}\int\frac{\rho\left(\textbf{r}'\right)}{\mid\textbf{r}-\textbf{r}'\mid}e^{-\xi\mid\textbf{r}-\textbf{r}'\mid}d^3r'
\label{3e}
\end{equation}
Here we have replaced $R^{(2)}$ by equation (\ref{3d}). So we see that in $f(R)$ gravity we have got Newtonian attractive term along with an attractive Yukawa term and also their strength gets modulated by the parameter $\delta$. In the limit $\delta\rightarrow0$ and $\xi\rightarrow\infty$ Newtonian gravity is obtained. Also in the limit $\xi\rightarrow\infty$ gravity becomes weakest whereas in the limit $\xi\rightarrow0$ gravity becomes strongest for any fixed $\delta$.

The effective potential and radial acceleration for a spherically symmetric object of radius $\mathcal{R}$ can hence be given by
\begin{eqnarray}
&\Phi(r)&=-\frac{4\pi G_N}{\left(1+\delta\right)r}\int_{0}^{r}r'^2\rho(r')dr'-\frac{4\pi G_N}{1+\delta}\int_r^{\mathcal{R}}r'\rho(r')dr'\nonumber \\&-&\frac{4\pi G_Ne^{-\xi r}}{3\left(1+\delta\right)\xi r}\int_{0}^{r}r'\rho(r')\sinh\left(\xi r'\right)dr'-\frac{4\pi G_N}{3\left(1+\delta\right)\xi r}\sinh\left(\xi r\right)\int_{r}^{\mathcal{R}}r'\rho(r')e^{-\xi r'}dr'
\end{eqnarray}
\label{3f}
and 
\begin{eqnarray}
a\left(r\right)&=&-\frac{d\Phi}{dr}\nonumber\\
&=&-\frac{4\pi G_N}{(1+\delta)r^2}\int_{0}^{r}r'^2\rho(r')dr'-\frac{4\pi G_N}{3(1+\delta)\xi r^2}\left(1+\xi r\right)e^{-\xi r}\int_{0}^{r}r'\rho(r')\sinh\left(\xi r'\right)dr'\nonumber \\&&-\frac{4\pi G_N}{3(1+\delta)\xi r^2}\left[\sinh\left(\xi r\right)-\xi r \cosh\left(\xi r\right)\right]\int_{r}^{\mathcal{R}}r'\rho(r')e^{-\xi r'}dr'
\label{fR}
\end{eqnarray}
\section{White dwarf stars as a probe of modified gravity}\label{n}

Hydrostatic equilibrium in a star is maintained as the radiation pressure emanating from the thermonuclear reactions occurring in the stellar interior balances the inward gravitational pull. But once the star exhausts its own nuclear fuel towards the late phase of its evolution, the core of the star starts contracting. Hence the density of the core starts increasing. Once the density goes above a certain threshold quantum mechanical effects start showing up, the core becomes degenerate and the envelope is expelled through different ejection mechanisms like shedding of outer shell as planetary nebula or a supernova explosion. The electrons become degenerate much before heavier particles like neutrons and the stellar configuration in which electron degeneracy pressure balances the inward gravitational pull is termed as a white dwarf. 

We now recall a simple model of carbon-oxygen white dwarfs. We assume the degenerate electron gas is in the ground state and the star is in a completely ionized state. We also neglect electrostatic corrections and general relativistic  effects in this study. On the basis of the above considerations, we derive the equation of state of relativistic electron gas. We follow the formalism developed in \cite{Shapiro} to discuss the physics of white dwarfs.

Since the electrons behave as an ideal Fermi gas and they are confined in the Fermi sphere at zero temperature, the number density of electrons can be obtained by computing the phase space integral of the Fermi distribution function which is essentially a step function at zero temperature over the Fermi sphere \cite{Huang}
\begin{equation}
n_e=\frac{m_e^3}{3\pi^2}x^3
\end{equation}     
where $x=\frac{p_F}{m_e}$, $p_F$ being the Fermi momentum. Since the carbon (oxygen atoms also have same number of electrons per nucleon, hence the analysis also holds for the oxygen in the same way) ions are not relativistic, their energy density is given by
\begin{equation}
\rho_c=\frac{n_e}{6}m_c=\frac{m_c m_e^3}{18\pi^2}x^3
\label{c}
\end{equation}
where $m_c$ and $m_e$ are the masses of carbon nuclei and electrons. 
The energy density of electrons can be obtained as \cite{Shapiro}
\begin{eqnarray}
\rho_e &=& 2\int_{0}^{p_F} \frac{\sqrt{p^2c^2+m^2c^4}}{8\pi^3}d^3p\nonumber \\
&=&\frac{m_e^4}{8\pi^2}\left[x\sqrt{1+x^2}\left(1+2x^2\right)-
\log_e\left(x+\sqrt{1+x^2}\right)\right]
\label{density}
\end{eqnarray}
Hence the total energy density $\rho$ is given by $\rho_c+\rho_e$ and it will be basically dominated by $\rho_c$ as $m_c \gg m_e$. The degeneracy pressure of the relativistic electrons can be computed from the standard expression in kinetic theory of gases \cite{Shapiro}
\begin{eqnarray}
P_e &=& \frac{1}{3}\int_{0}^{p_F}v p f(p)4 \pi p^2 dp\nonumber\\
&=&\frac{m_e^4}{8 \pi^2 }\left[x\sqrt{1+x^2}\left(\frac{2}{3}x^2-1\right)+
\log_e\left(x+\sqrt{1+x^2}\right)\right]
\label{pressure}
\end{eqnarray}
Any contribution to the pressure coming from carbon-oxygen ions is neglected in this study. So we have established a relation between density and degeneracy pressure through $x$. In the ultra-relativistic ($x \gg 1$) and non-relativistic ($x \ll 1$) limits, this relation takes a simple form $P=K \rho^{1+\frac{1}{n}}$ where $K$ is a constant and $n$ is the index which for the two mentioned cases takes the value $3$ and $\frac{3}{2}$  respectively \cite{Shapiro}. 
 
Let us now consider the stellar structure equations for white dwarfs. The mass continuity equation is given by
\begin{equation}
\frac{dm\left(r\right)}{dr}=4\pi r^2 \rho(r)
\label{mass}
\end{equation}
The momentum conservation equation i.e. Euler equation reads as
\begin{equation}
\frac{dP_e\left(r\right)}{dr}=\rho(r)a(r)
\end{equation}
By using the expression for degeneracy pressure (\ref{pressure}) the above relation can be written as
\begin{equation}
\frac{dx}{dr}=\frac{\sqrt{1+x^2}}{x^4}\frac{3\pi^2}{m_e^4}\rho(r)a(r)
\label{x}
\end{equation}
Since $\rho(r)$ can also be expressed as a function of $x(r)$, the equations (\ref{mass}) and (\ref{x}) form a set of coupled first order differential equations for the functions $m(r)$ and $x(r)$. Gravity enters into the problem through the acceleration term $a(r)$ which for Newtonian gravity or Einstein gravity in the Newtonian limit takes the form $\frac{-Gm(r)}{r^2}$; however, for modified gravity we get additional terms as we have seen in the preceding sections. Hence the white dwarfs exhibit different signatures for different modified gravity theories and one can constrain these theories from white dwarf observations.   
\section{Results and Discussion}\label{o}
In the previous section we have obtained the stellar structure equations 
(\ref{mass}, \ref{x}) for studying the physics of white dwarfs. We will solve these coupled first order differential equations numerically, using a standard RK4 method with the initial conditions $m(0)=0$ and $x(0)=x_0$ where $x_0$ is related to the central density of the white dwarf through the equations (\ref{c}) and (\ref{density}). The mass of the white dwarf is defined as $M=m(\mathcal{R})$ where $\mathcal{R}$, the radius of the star, corresponds to that value of $r$ at which pressure or $x$ goes to zero. For Newtonian gravity, the mass and radius of the star essentially depend upon the value of $x_0$. But in case of modified gravity theories the mass and radius depend on $x_0$ as well as the parameter values of those theories. In case of STVG, FOG and $f(R)$ gravity the emergence of non-local terms in the expression for acceleration [equations (\ref{STVG}), (\ref{FOG}) and (\ref{fR})] i.e. the terms with integral from $r$ to $\mathcal{R}$ make the analysis complicated as one does not expect to know the information about that density profile of white dwarfs a priori. Hence we first obtain the density profile for the entire star assuming Newtonian gravity. Then we feed that profile into those non-local integrals to obtain the initial mass and radius of the star. We use these initial values to obtain the mass and radius for the next iteration and we continue this process iteratively until we get a precision of mass of the star of order of $10^{-4}$. We have used standard Trapezoidal rule for computing the integrals. To check the accuracy of the results, we have also used Simpson's rule for computing the integrals and found no difference in results up to a very high precision.
\subsection{Mass-Radius relation}
We explore the mass radius relationship of the white dwarfs for the modified gravity theories discussed in (\ref{m}). In this study, we have restricted the upper value of $x_0$ to $27$ which corresponds roughly to the central density $3.9\times10^{10}$ gm/cc to avoid possible neutronization for carbon-oxygen white dwarfs \cite{Shapiro}. In case of Newtonian gravity the maximum mass of white dwarf that we have found is $\sim 1.44 M_{\odot}$ which is the Chandrasekhar limit \cite{Chandra}. Obviously, for modified gravity models this limit would either be enhanced or reduced because of additional attractive or repulsive terms in the expression for acceleration. Also the magnitude of enhancement or reduction of the maximum mass limit would depend upon the parameters of the modified gravity model.
\subsubsection{\textbf{STVG}}
\begin{figure}[h]
\includegraphics[width=0.49\textwidth]{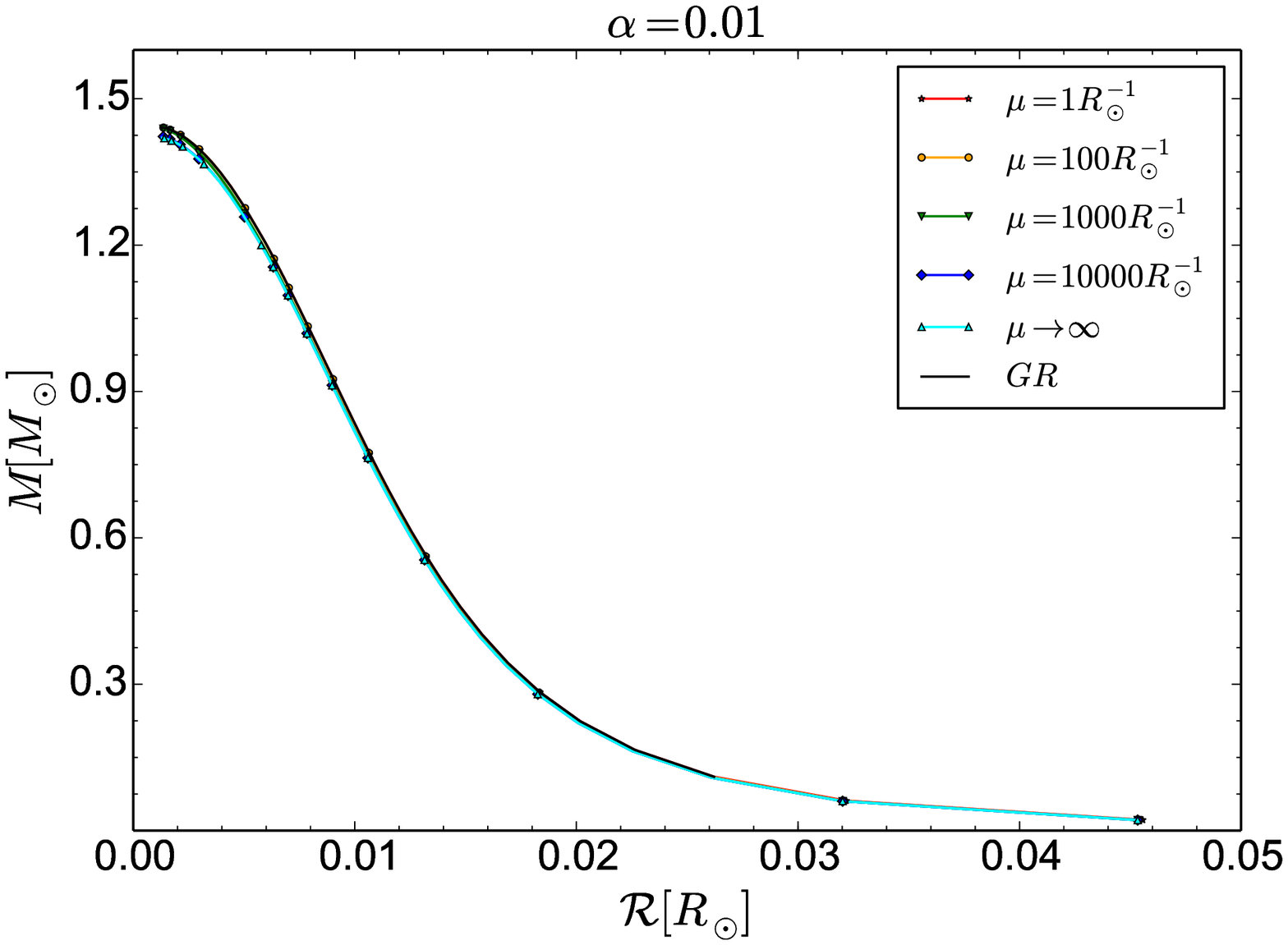}
\includegraphics[width=0.49\textwidth]{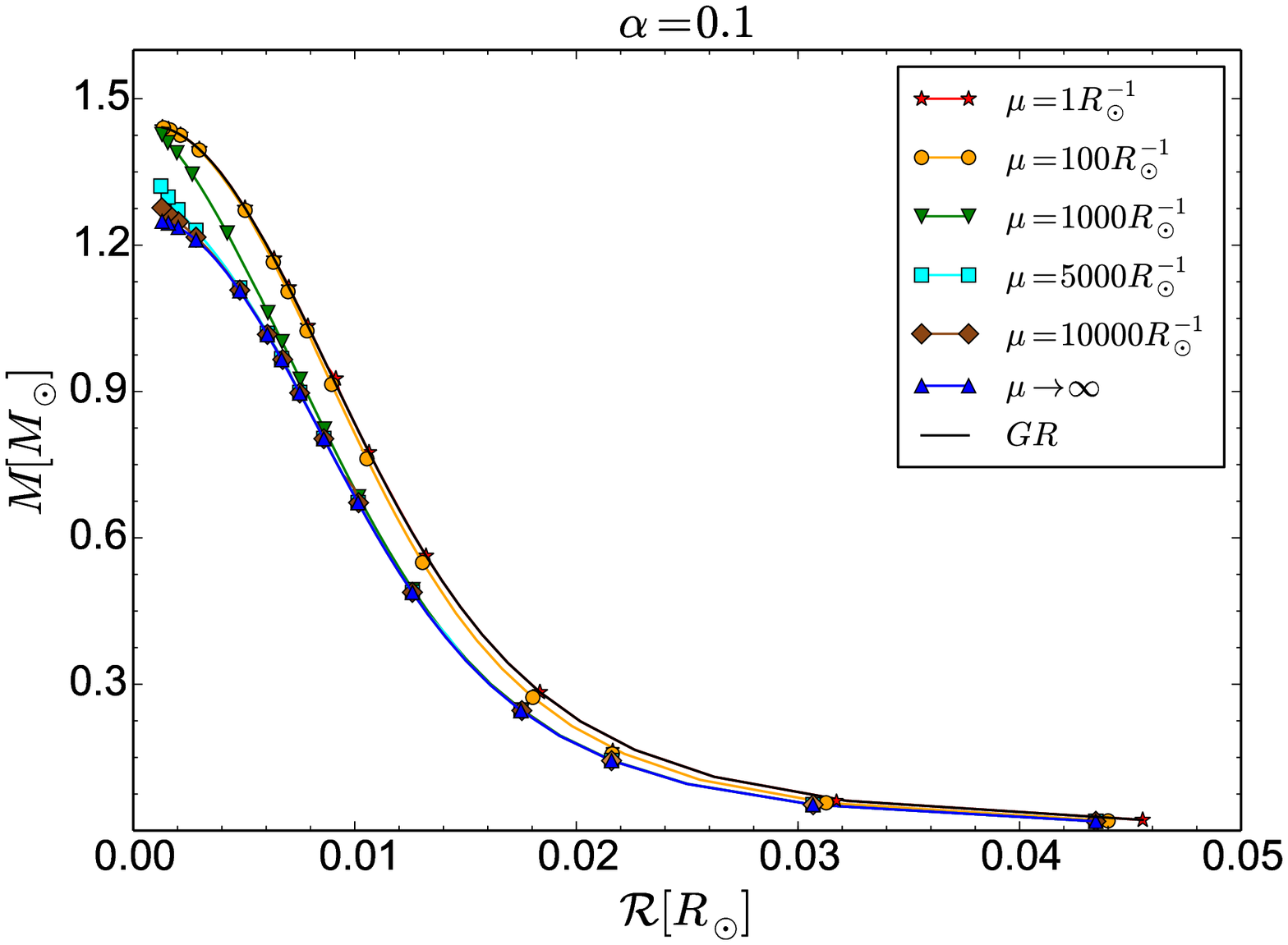}
\includegraphics[width=0.49\textwidth]{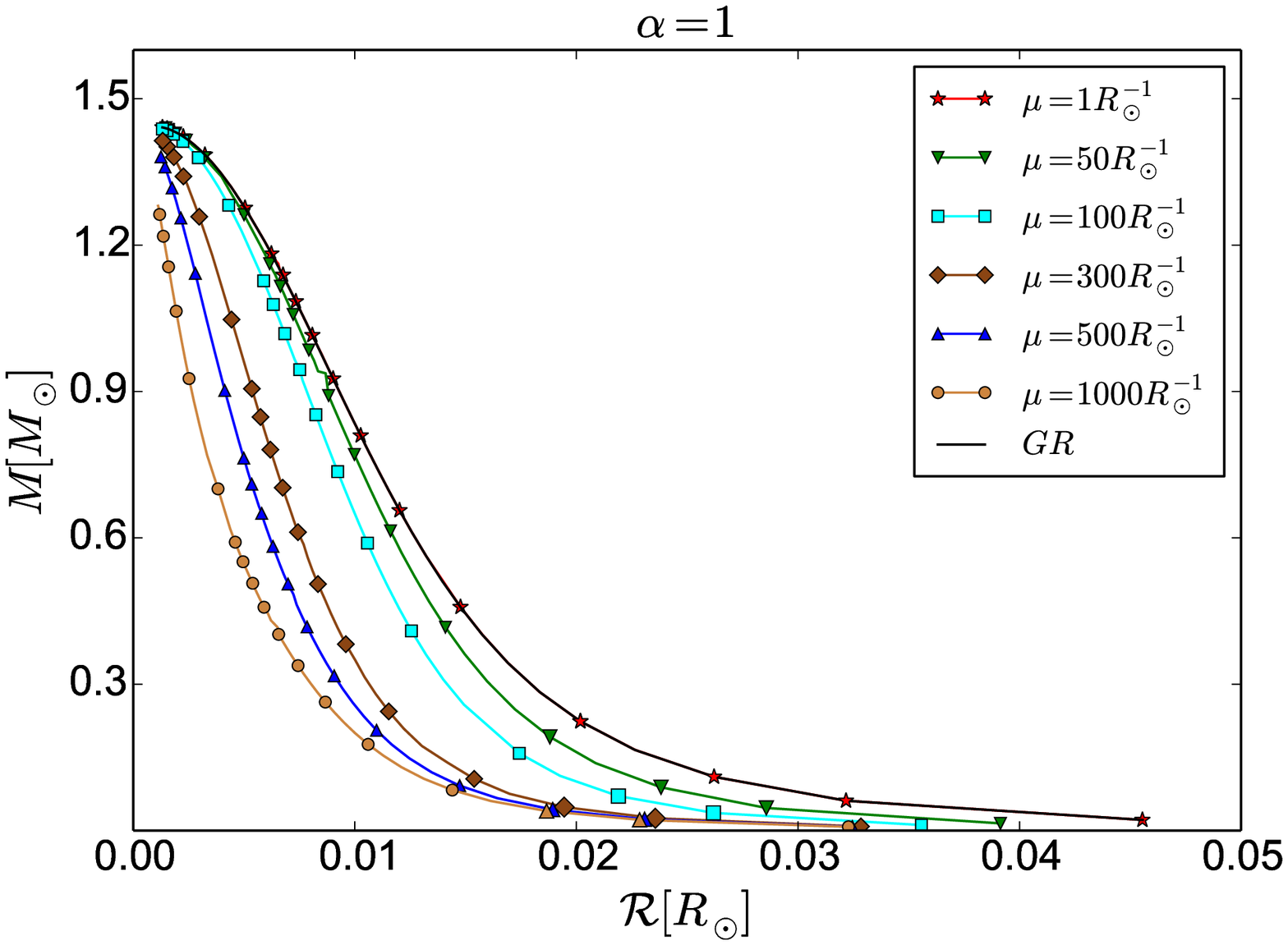}
\includegraphics[width=0.49\textwidth]{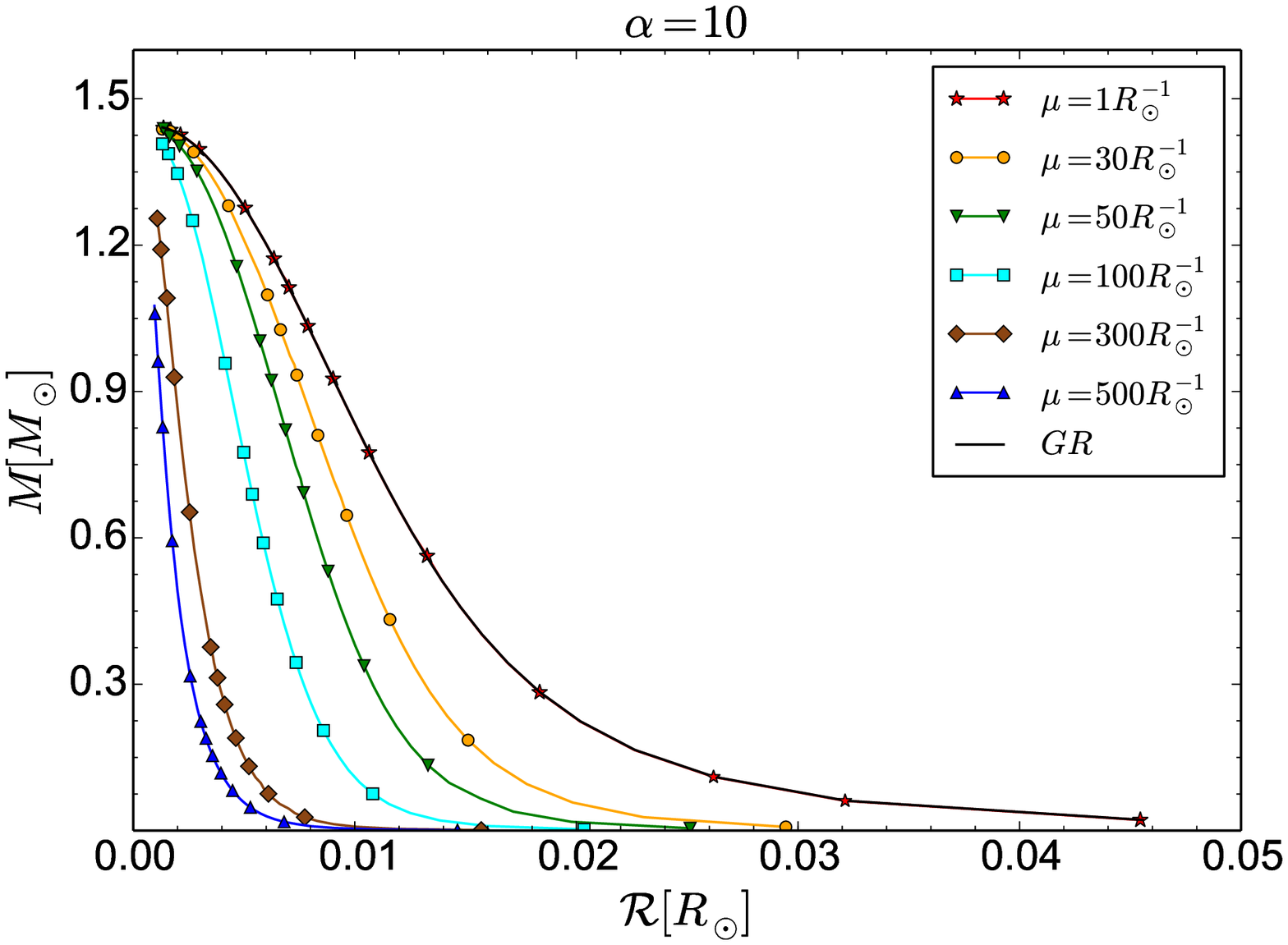}
\begin{center}
\includegraphics[width=0.49\textwidth]{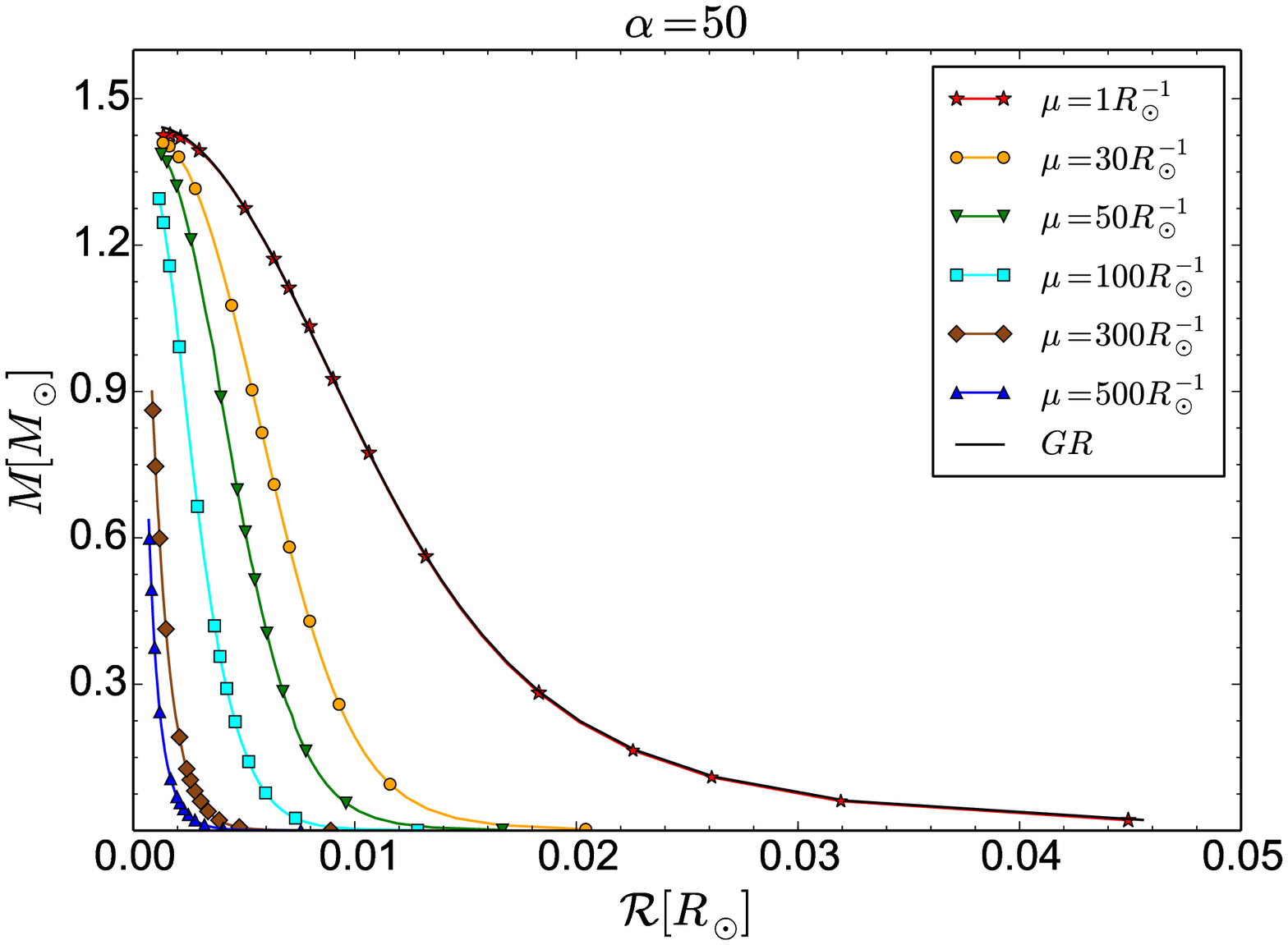}
\end{center}
\caption{The total mass $M$ of white dwarf star in STVG against its radius $\mathcal{R}$ for various values of the parameters $\alpha$ and $\mu$. In a given plot, we have fixed $\delta$ and $M-\mathcal{R}$ curves are obtained for various values of $\mu$.}
\label{fig1}
\end{figure}

As we have seen in the section (\ref{stvg}), STVG is described by two parameters $\alpha$ and $\mu$ and its effective potential contains a Yukawa repulsive term as well as the enhanced Newtonian attractive term (\ref{1f}). But as the repulsive term is always smaller than the attractive term for $\alpha>0$, gravity in STVG is stronger than  Newtonian gravity. Hence Chandrasekhar mass limit will decrease compared to that of Newtonian gravity (See FIG. \ref{fig1}).

As one can see from the expression (\ref{1f}), Newtonian gravity is recovered in the limit $\alpha\rightarrow0$ or $\mu\rightarrow0$. Also, gravity becomes strongest in this model in the limit $\mu\rightarrow \infty$. In between these two limits the structure of the white dwarfs is basically controlled by the exponential factor in Yukawa term along with $\alpha$.
For any fixed $\alpha$, the repulsive term weakens with increasing $\mu$ and hence the mass limit decreases (See FIG. \ref{fig1}). Similarly, for fixed $\mu$, gravity becomes more attractive as $\alpha$ increases and hence also the Chandrasekhar limit decreases - this is also captured in  FIG.(\ref{fig1}). 
\subsubsection{\textbf{EiBI}}
\begin{figure}[ht]
{
\includegraphics[width=0.49\textwidth]{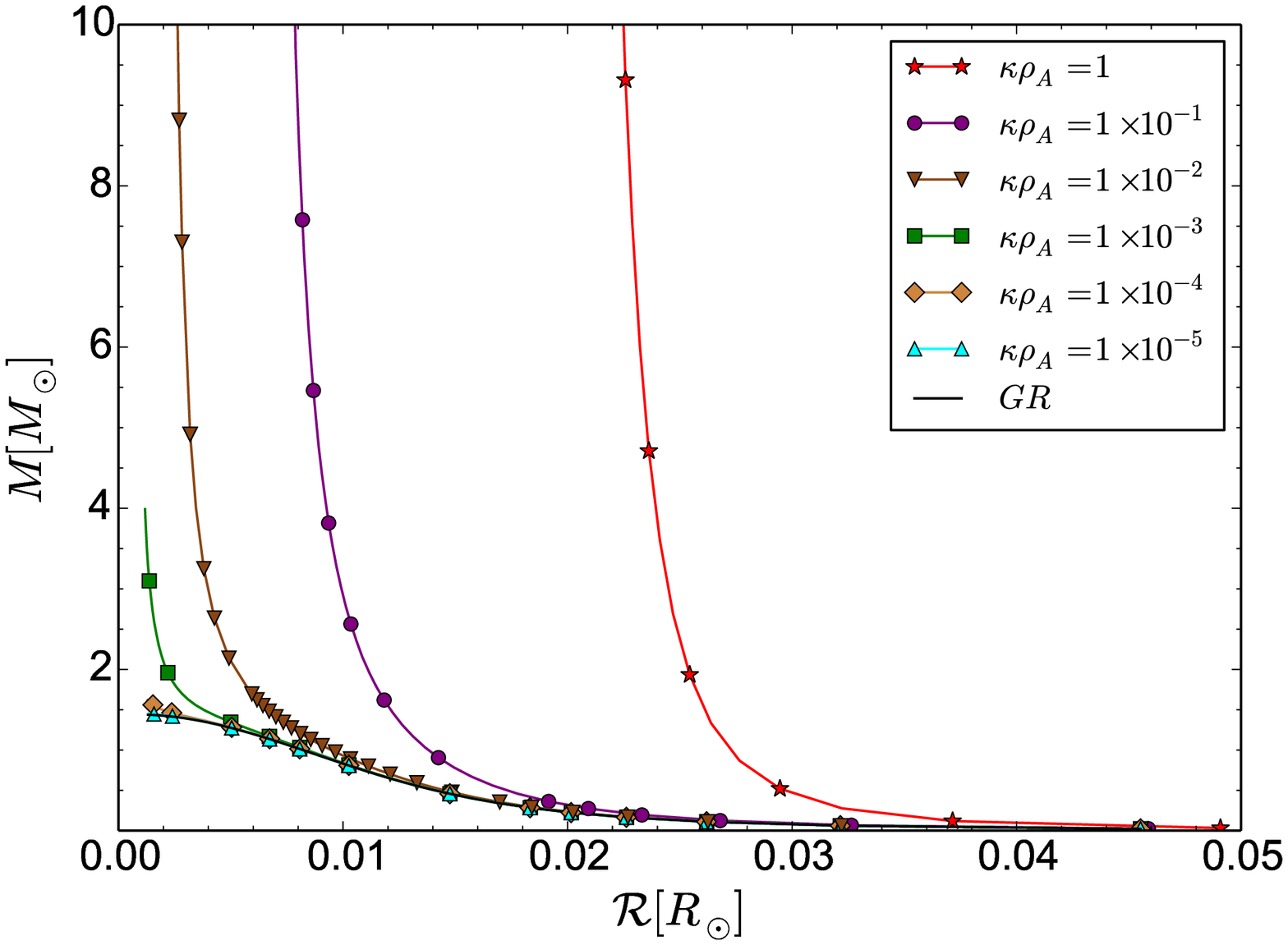}
\includegraphics[width=0.49\textwidth]{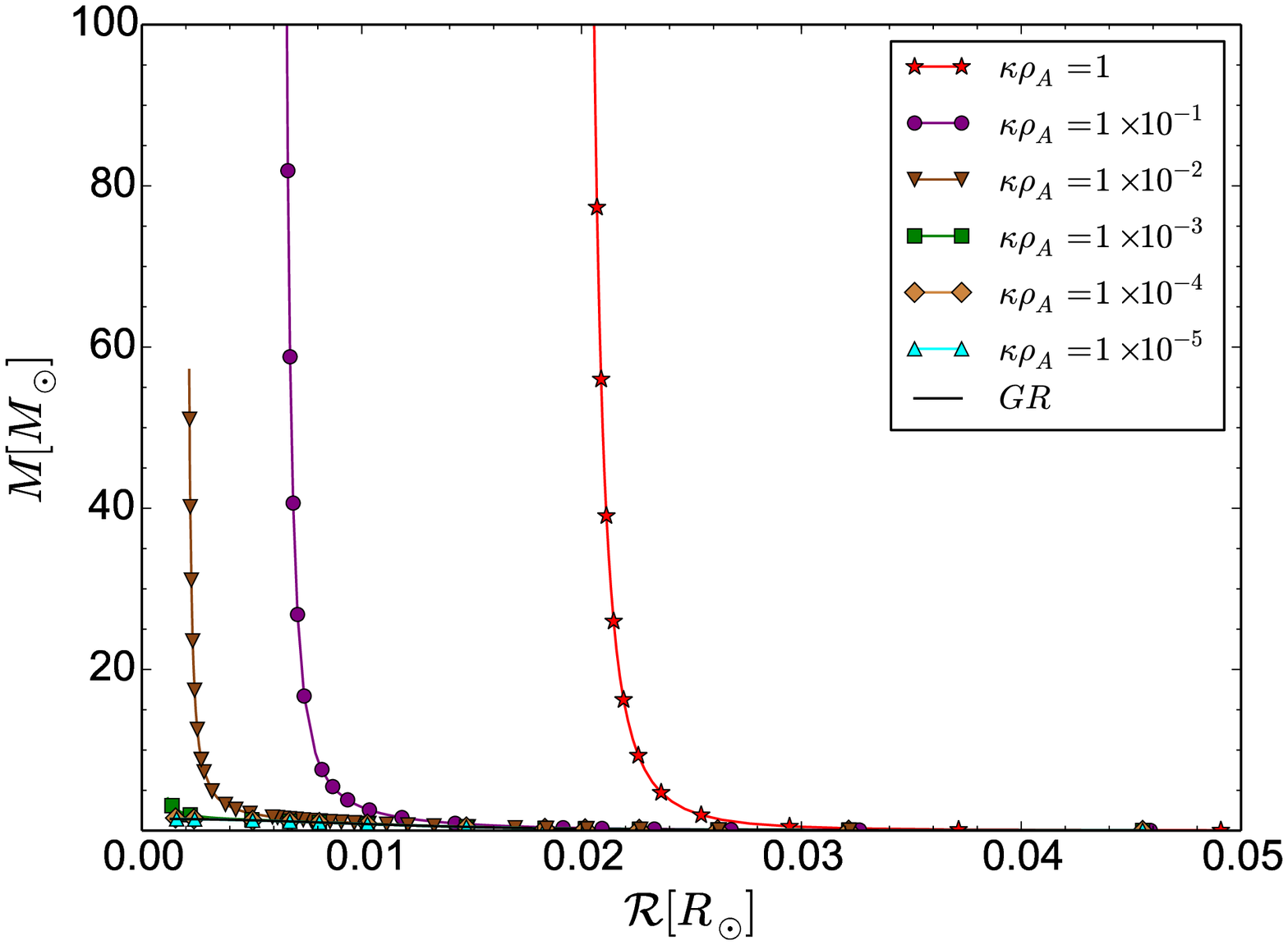}
\caption{Total mass $M$ of the white dwarfs in EiBI gravity ($\kappa>0$) against the radius $\mathcal{R}$ is plotted for different values of the parameter $\kappa(>0)$. Here $\rho_A=\rho_{\odot}\times10^9$. In both the plots same M-$\mathcal{R}$ diagrams are shown except that the maximum scale value of $M$ is different in two plots highlighting the feature that mass does not stabilize to a particular value.}
\label{fig2a}
}
\end{figure}

\begin{figure}
{
\begin{center}
\includegraphics[width=0.49\textwidth]{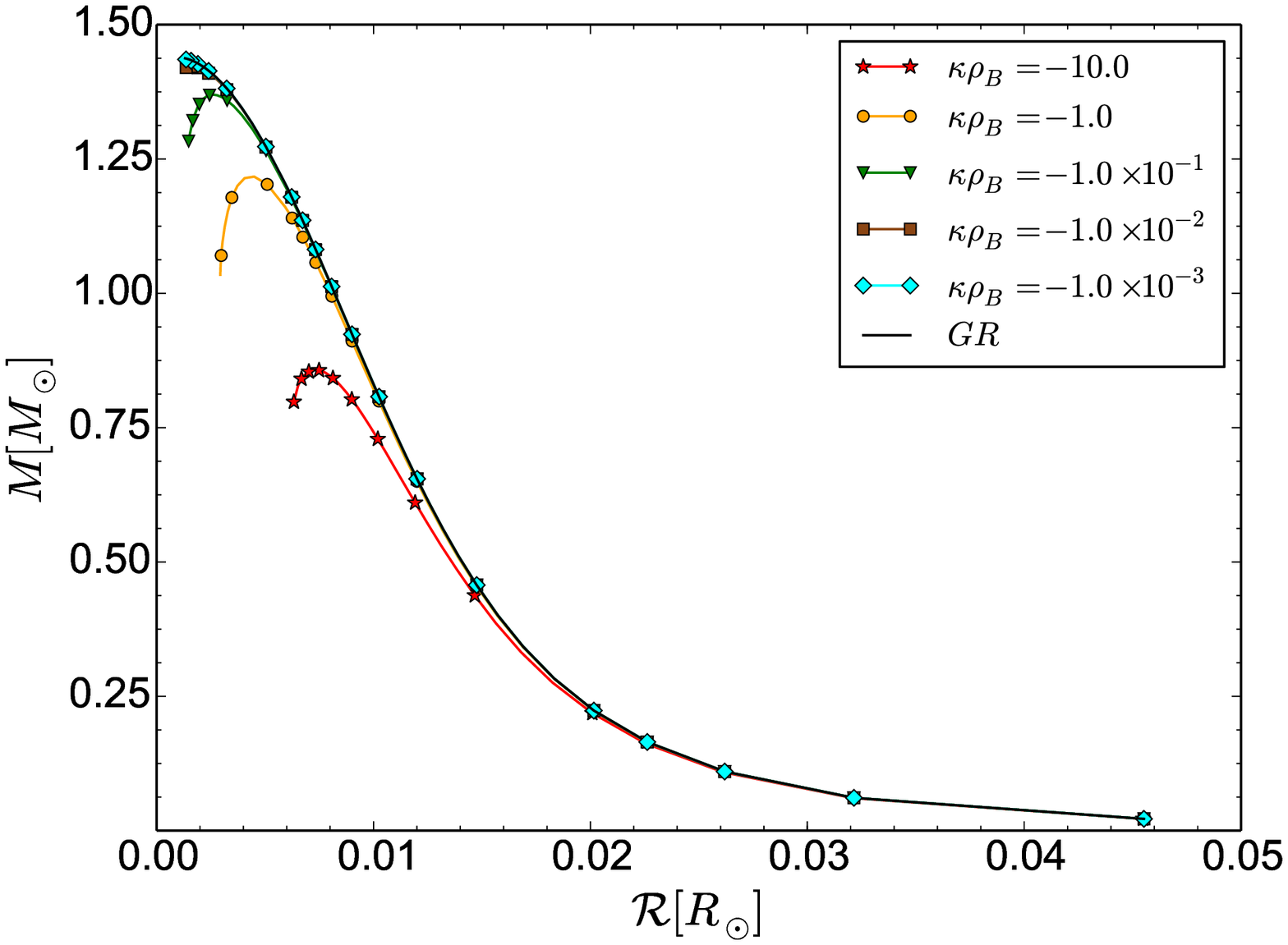}
\caption{Total mass $M$ of the white dwarfs in EiBI gravity ($\kappa<0$) against the radius $\mathcal{R}$ is plotted for different values of the parameter $\kappa (<0)$. Here $\rho_B=\rho_{\odot}\times10^{12}$.}
\label{fig2b}
\end{center}
}
\end{figure}

EiBI gravity (Section \ref{eibi}) has got only one parameter $\kappa$ and it can take positive as well as negative values. We would consider both the cases. When $\kappa>0$ the additional part in the acceleration behaves as a repulsive force (See equation \ref{EiBI}) and hence white dwarfs would be able to support more mass depending upon the value of the parameter (FIG. \ref{fig2a}). The higher the parameter value, the higher the mass it can support (FIG. \ref{fig2a}). For negative values of $\kappa$ we would obviously see the opposite features i.e. gravity would become stronger than Newtonian case and hence white dwarfs would be less massive (FIG. \ref{fig2b}). The $\kappa>0$ case has been considered in \cite{Pani} and they have found that Chandrasekhar limit does not exist for this case i.e. the mass does not stabilize to a particular value even for very high $x$. Rather it would go on increasing as one increases $x$ and it can have very high value ($>100M_{\odot}$) depending upon the value of $\kappa$. We have also got the same behaviour (FIG. \ref{fig2a}) and also it seems there exists a critical radius $\sim \sqrt{\kappa}$ below which a white dwarf cannot exist similar to what \cite{Pani} have obtained. For negative values of $\kappa$ as mentioned above mass limit decreases from the Chandrasekhar limit in Newtonian case and below a particular value (it depends upon the central density) no stable white dwarfs exist as gravity would be so strong that electron degeneracy pressure would not be able to support gravity. This has been discussed in \cite{Pani} for any polytropic model of the form $P=K\rho^{1+\frac{1}{n}}$ ($n$ being the polytropic index). They have shown $\kappa$ has to be greater than $-4K(1+\frac{1}{n})\rho_{0}^{-1+\frac{1}{n}}$ ($\rho_0$ is the central density) otherwise the stellar object would not exist. For $x_0=0.1$, the above condition gives $k>-5.3\times10^5 m^5 kg^{-1}s^{-2}$.

Recent discoveries of several highly over-luminous Type Ia supernovae (SNe Ia) like SN 2003fg, SN 2007if, SN 2009dc have suggested that their progenitor white dwarfs may have 
mass in the range 2.1-2.8$M_{\odot}$ implying the existence of super-Chandrasekhar white dwarfs \cite{Hicken,Howell,Scalzo,Silverman,Taubenberger}. Several theories have been proposed to explain their existence \cite{Bani,Chatterjee,Hachisu,Bhatta} and here we consider the perspective of modified gravity. Since for positive values of $\kappa$ white dwarfs can have mass more than the usual Newtonian Chandrasekhar limit, one can effectively constrain the parameter space of $\kappa$ by considering the maximum mass that a white dwarf can have as $2.8M_{\odot}$ which is the estimated mass of the progenitor carbon-oxygen white dwarf for SN 2009dc \cite{Taubenberger}. It gives a reasonable constraint $\kappa<0.35\times10^{2}$ m$^5$ kg$^{-1}$s$^{-2}$. The most stringent constraint on positive $\kappa$ comes from neutron stars \cite{Pani} in which one obtains $\kappa<10^{-2}$ m$^5$ kg$^{-1}$s$^{-2}$. The relevance of modified gravity in the context of super Chandrasekhar white dwarfs has previously been discussed in \cite{Banim}.   
\begin{figure}[h]
\begin{center}
\includegraphics[width=0.49\textwidth]{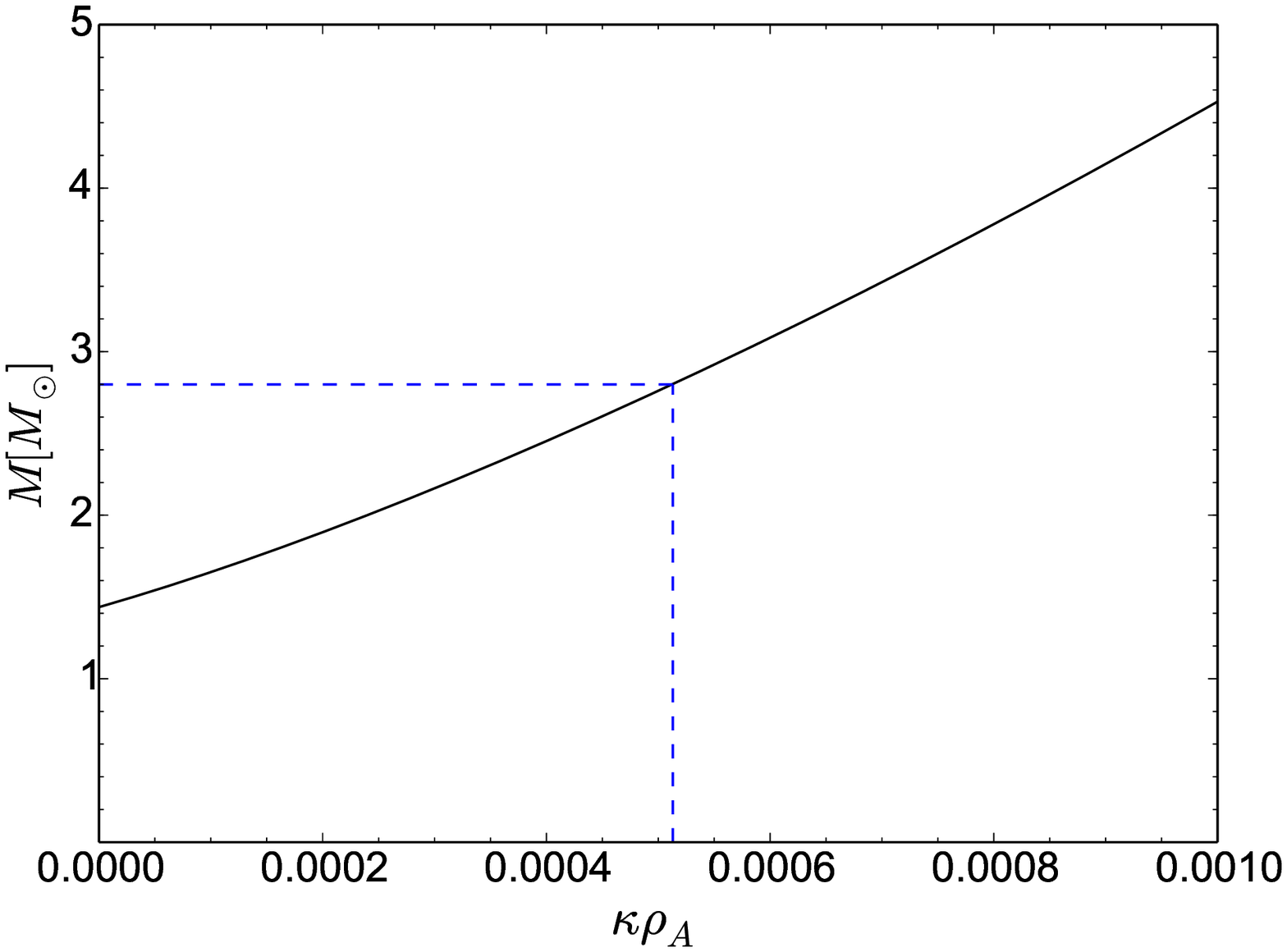}
\caption{Maximum mass of the white dwarf in EiBI gravity against the parameter $\kappa (>0)$. Here $\rho_A=\rho_{\odot}\times10^9$.}
\end{center}
\end{figure}
\subsubsection{\textbf{FOG}}
\begin{figure}[h]
\includegraphics[width=0.49\textwidth]{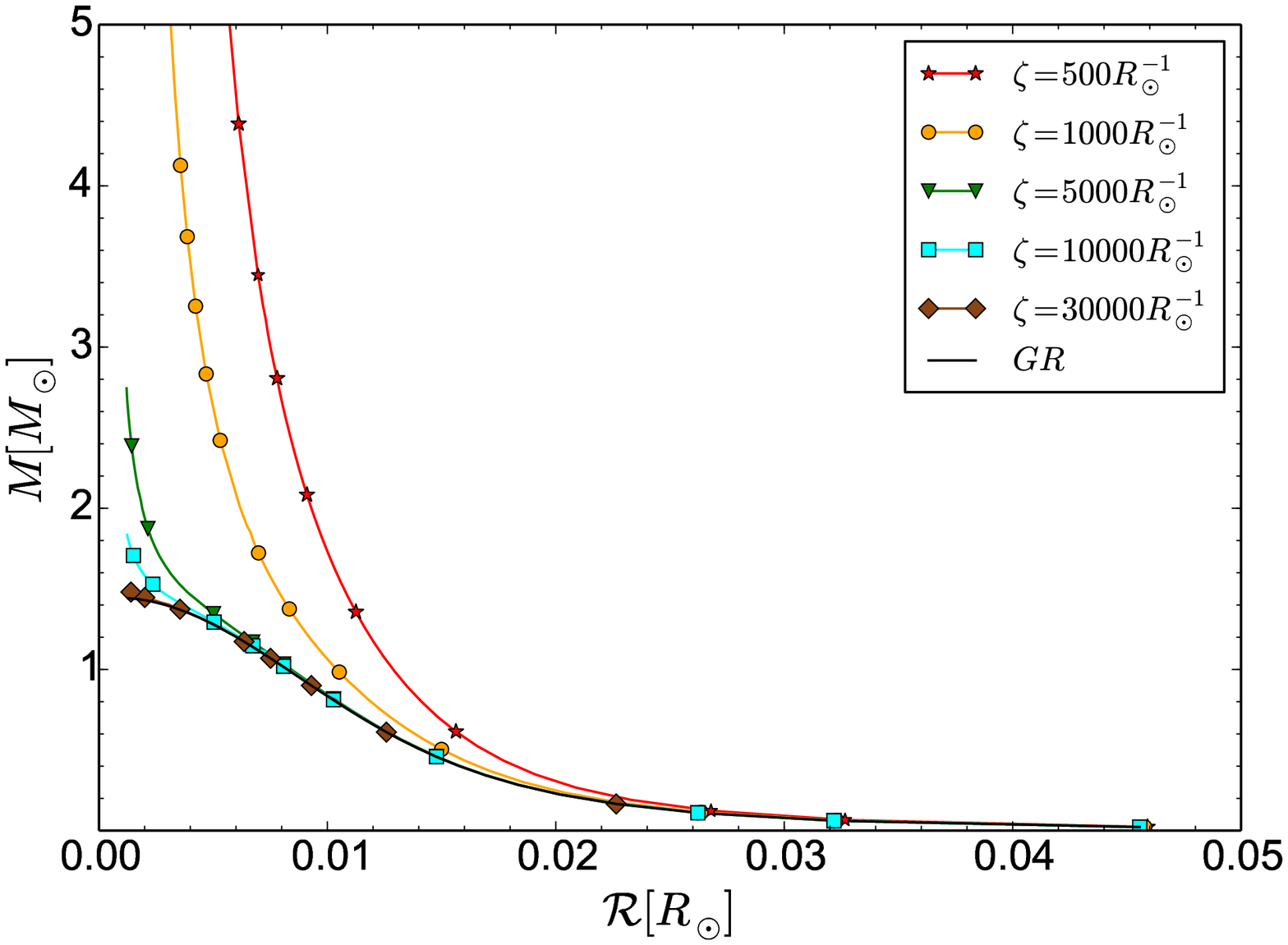}
\includegraphics[width=0.49\textwidth]{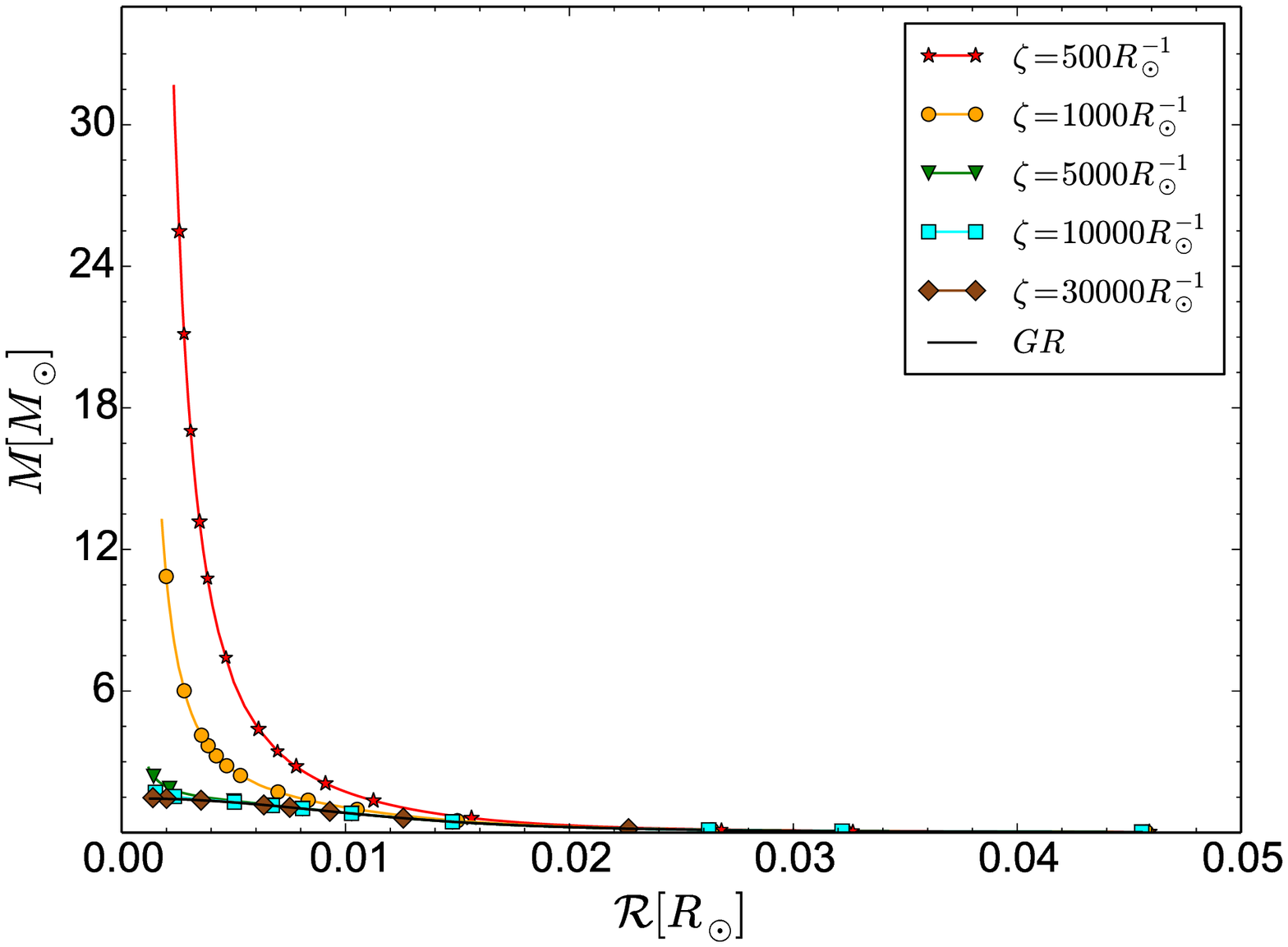}
\caption{Total mass $M$ of the white dwarfs in FOG model against its radius $\mathcal{R}$ is plotted for different values of the parameter $\zeta$.}
\label{fig3}
\end{figure}
\begin{figure}[h!]
\begin{center}
\includegraphics[width=0.49\textwidth]{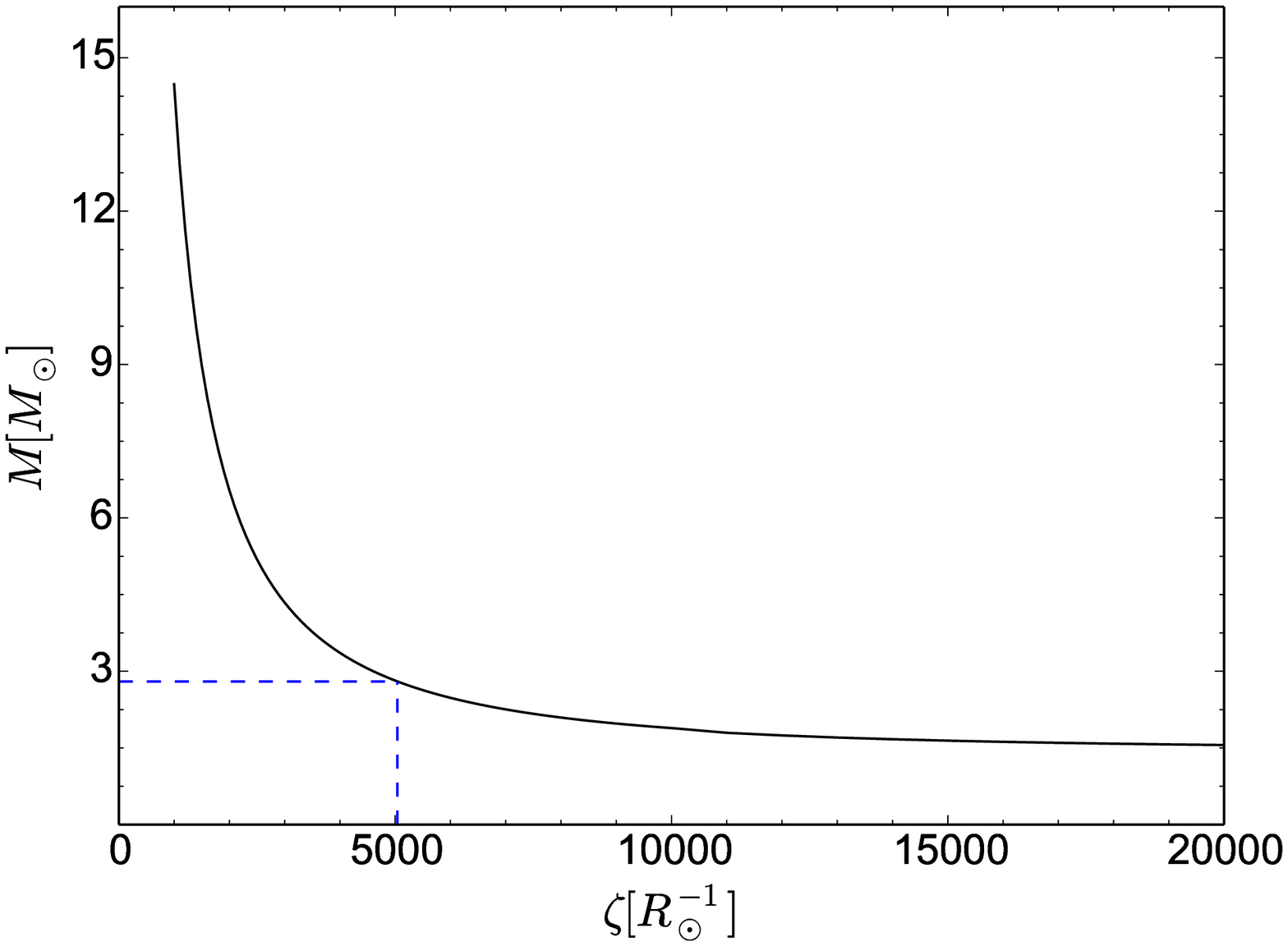}
\caption{Maximum mass $M$ of the white dwarf in FOG model against the parameter $\zeta$. The dotted line corresponds to $M$=$2.8 M_{\odot}$.}
\label{fig4}
\end{center}
\end{figure}
The fourth order gravity model discussed in Section \ref{fog} has one independent parameter $\zeta$ and it has got Yukawa repulsive term along with the usual Newtonian attractive term in the expression for effective potential (See Eqns. \ref{fog3}, \ref{FOG}). Therefore in this case also gravity would be weakened due to the presence of Yukawa repulsive term and it would be able to support more mass compared to the Newtonian case. As we see from the expression of effective potential (Eqn. \ref{fog3}), this model would converge to the Newtonian results in the limit $\zeta\rightarrow\infty$ and the deviation would increase as $\zeta$ decreases. The mass of the white dwarf in this model does not stabilize to a particular value and it goes on increasing with increasing $x$ like EiBI gravity (See FIG.\ref{fig3}). One can see from FIG. \ref{fig3} that this model also proposes a minimum radius for white dwarf depending upon the value of $\zeta$. Therefore one can constrain the parameter space of $\zeta$ by considering the fact that the maximum mass a super Chandrasekhar white dwarf can have is $2.8M_{\odot}$ as mentioned before and it poses a constraint $\zeta>4800 R_{\odot}^{-1}$ or $L(=\frac{1}{\zeta})<1.45\times10^5 m$ (See FIG. \ref{fig4}). This particular model in the context of neutron stars has been discussed in \cite{Santos} for a single equation of state of ideal neutron gas and a specific choice of the parameter $L=1.36\times 10^3m$. It was shown that stable neutron stars can exist even for arbitrarily large baryon numbers for the above mentioned equation of state and parameter value.
\subsubsection{$\textbf{f(R)}$}\label{mrf}
As we have seen in Section \ref{fr}, the $f(R)$ gravity model is characterized by two parameters $\xi$ and $\delta$. It has got one attractive Yukawa term along with the Newtonian attractive term, both of them being modulated by the parameter $\delta$. Unlike the FOG or EiBI gravity, the maximum mass of a white dwarf in $f(R)$ gravity model stabilizes to a particular value giving the Chandrasekhar limit for any fixed value of parameters. Gravity in this model can be stronger as well as weaker than the Newtonian gravity depending upon the parameter regime. In order to appreciate the mass-radius diagrams (FIG. \ref{fig5}), we have divided our parameter space into three regimes as given below.
\begin{enumerate}
\item $\xi\rightarrow\infty$ : Gravity becomes weakest in this limit for any fixed $\delta$ and it converges to the Newtonian result for $\delta=0$. $\delta>0$ weakens the gravity further resulting in an increase in Chandrasekhar mass limit. Also, the more one increases $\delta$, more the Chandrasekhar limit increases.
\item $\xi=0$ : Gravity becomes strongest in this model for any fixed $\delta$. In this case $\delta=\frac{1}{3}$ produces the Newtonian result. Below this value, gravity is stronger than Newtonian case, hence Chandrasekhar limit would decrease and vice-versa.
\item $\infty>\xi>0$ : Gravity is either stronger or weaker compared to Newtonian case depending upon the parameter values. But as $\xi$ increases keeping $\delta$ fixed, gravity becomes weaker and hence Chandrasekhar limit increases. Also as $\delta$ increases keeping $\xi$ fixed, gravity again becomes weaker resulting in an increase in Chandrasekhar limit.
\begin{figure}[h!]
\includegraphics[width=0.49\textwidth]{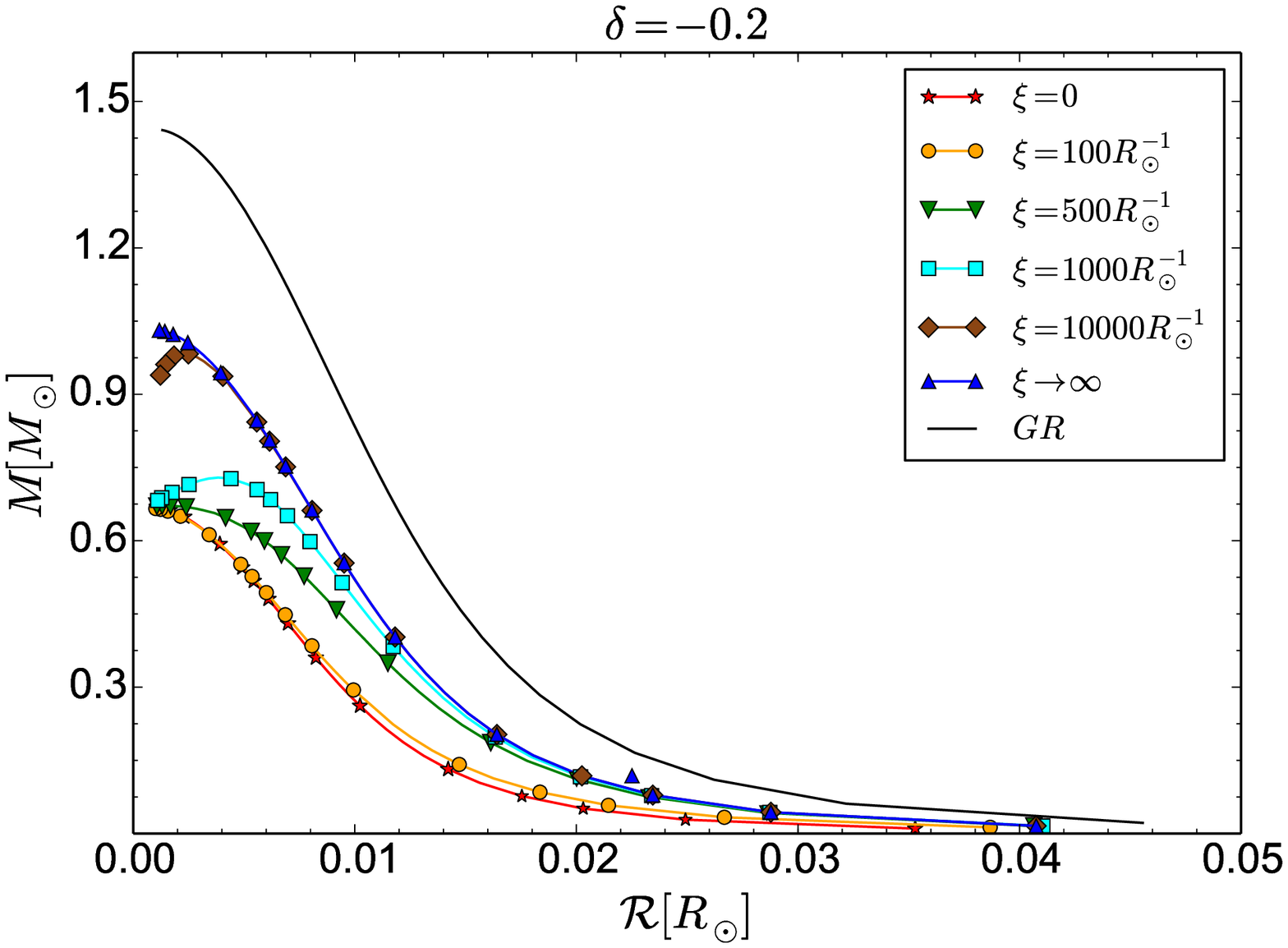}
\includegraphics[width=0.49\textwidth]{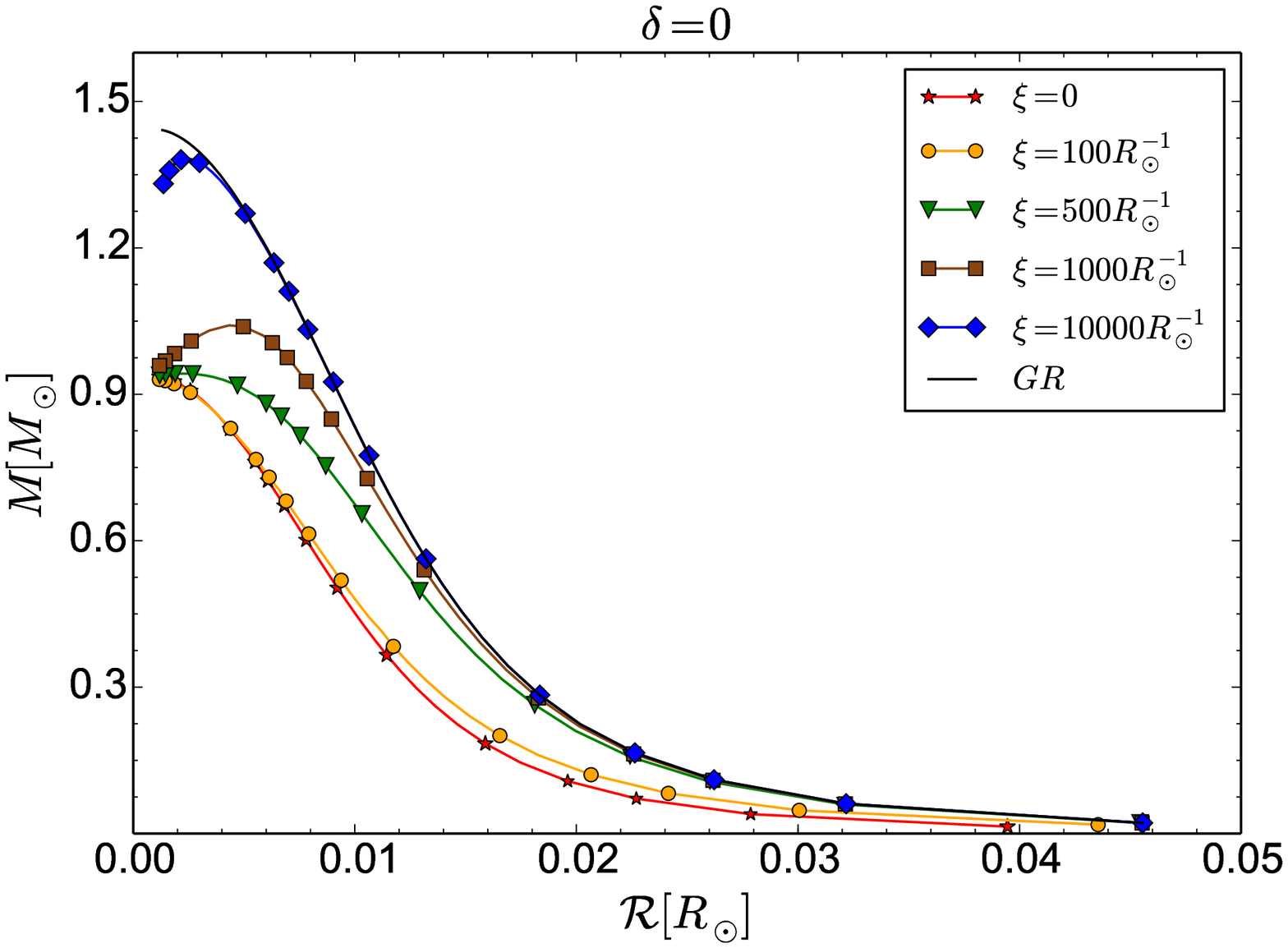}
\includegraphics[width=0.49\textwidth]{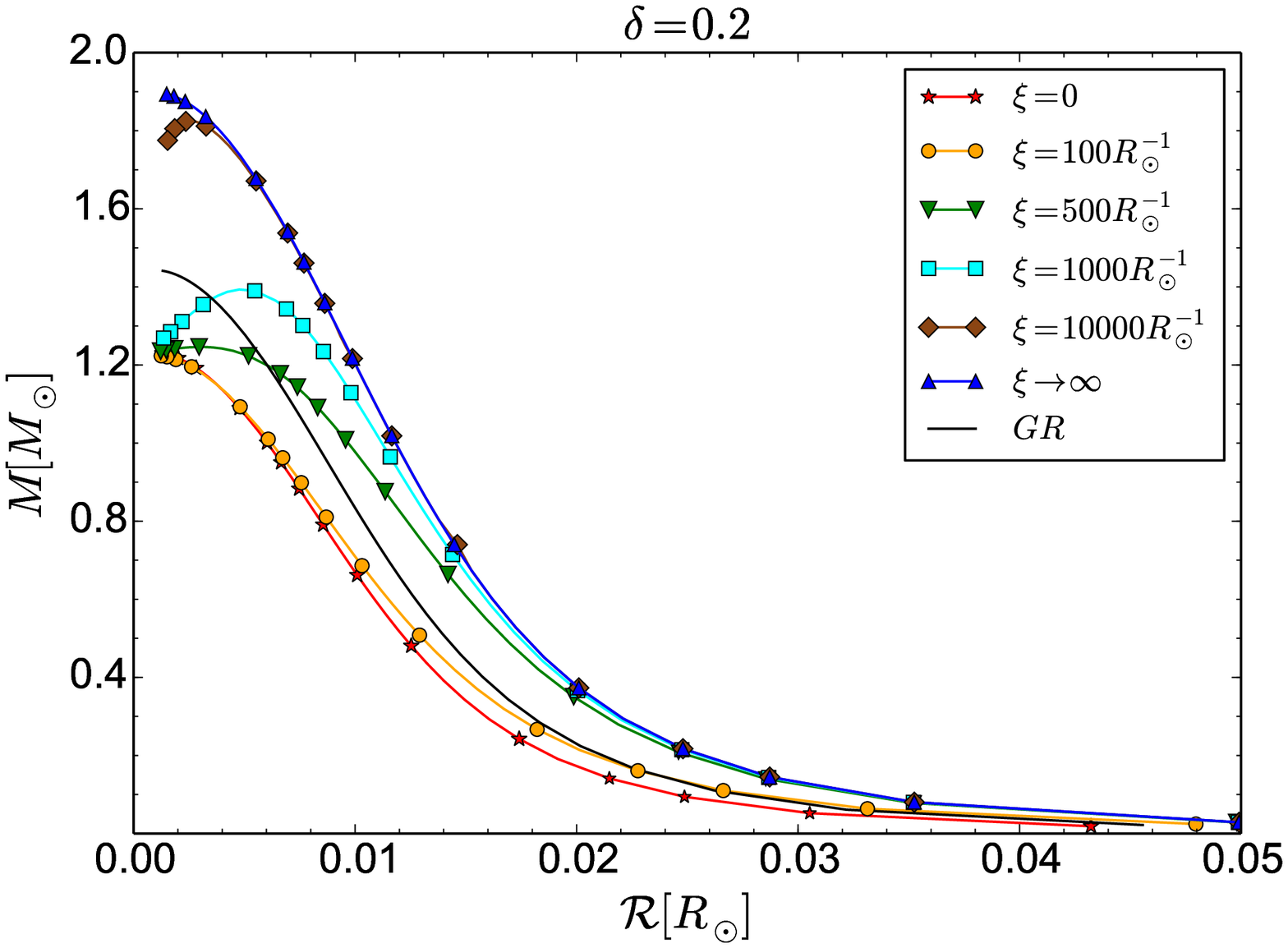}
\includegraphics[width=0.49\textwidth]{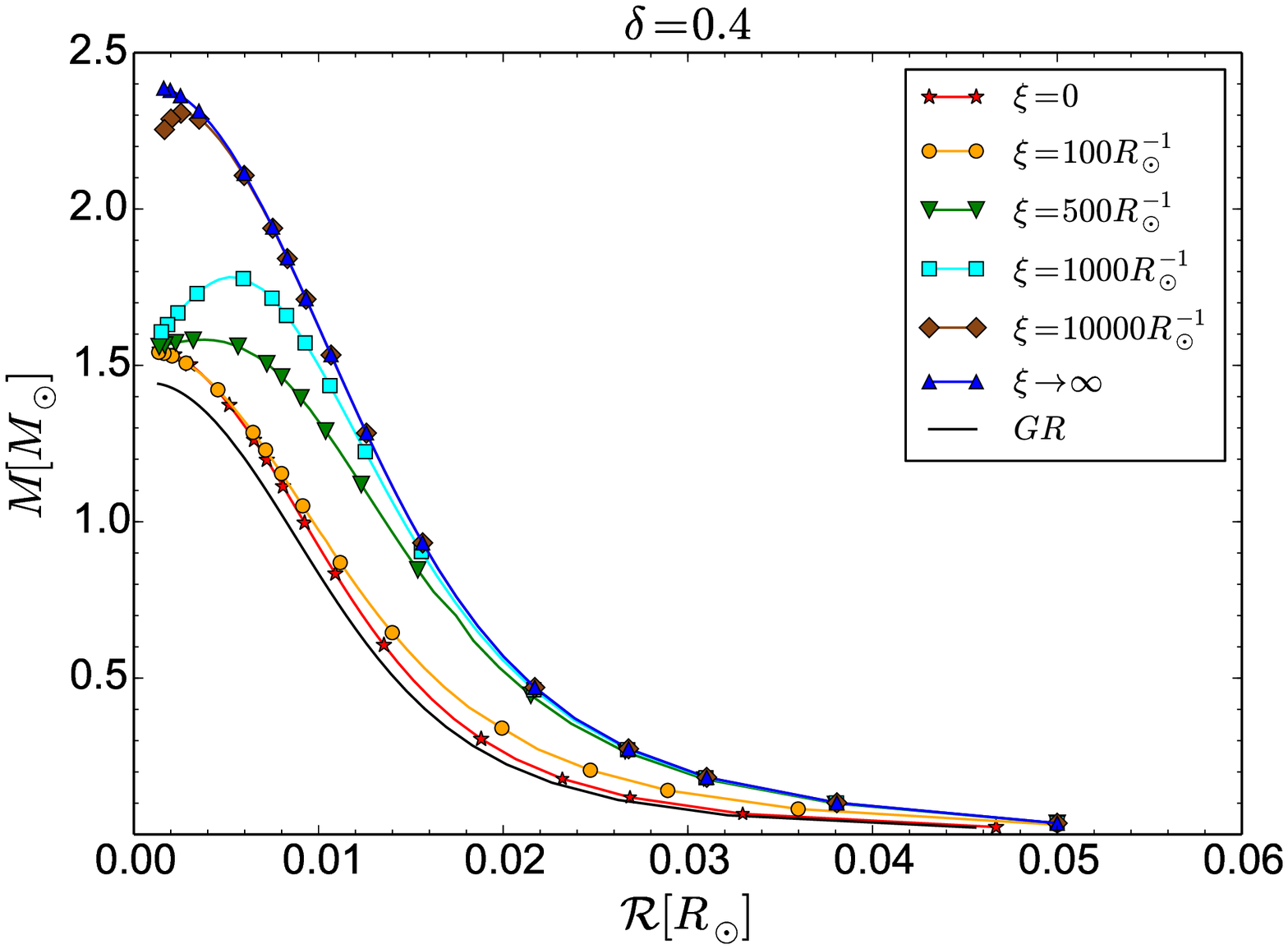}
\begin{center}
\includegraphics[width=0.49\textwidth]{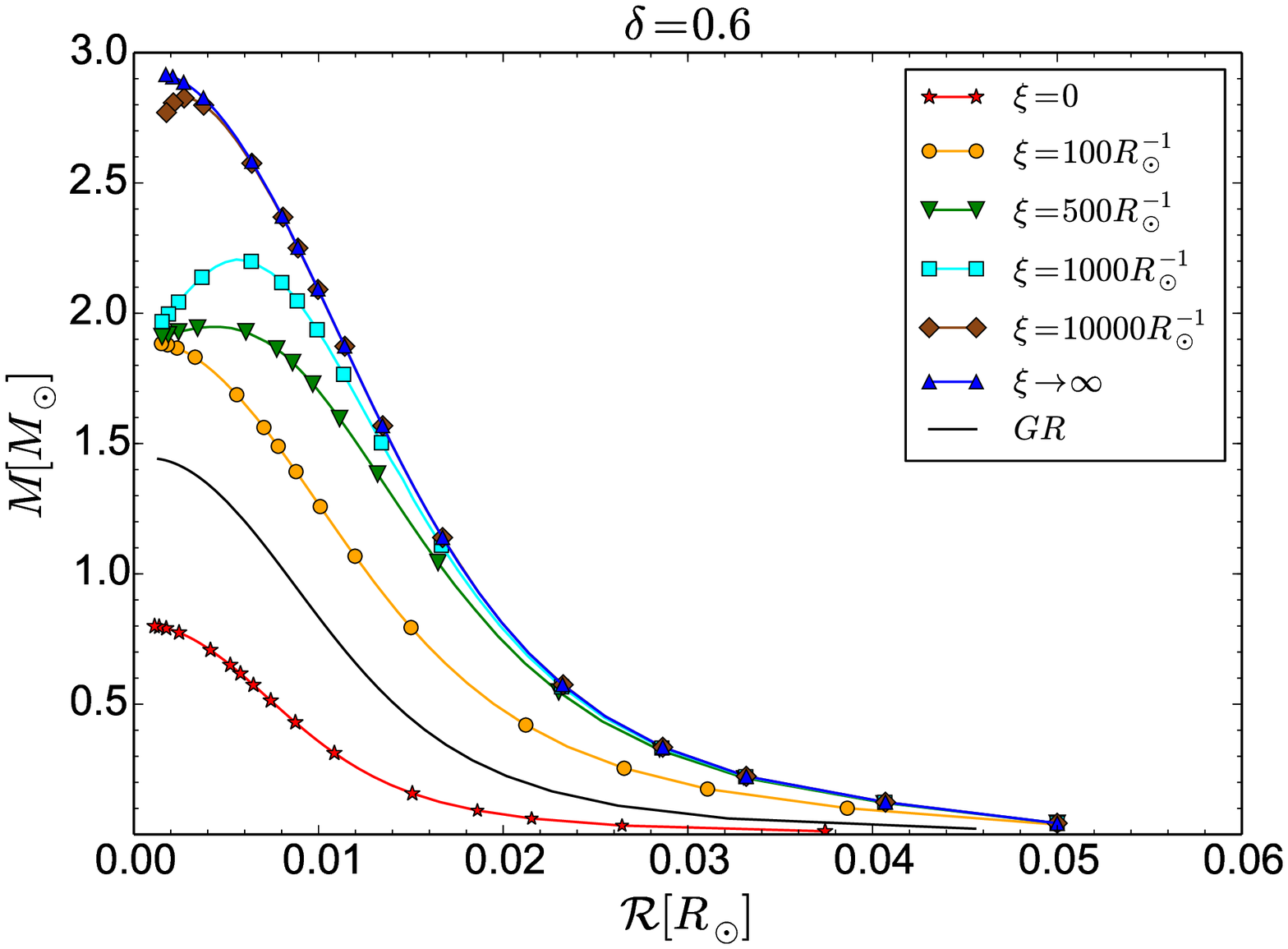}
\end{center}
\caption{Total mass $M$ of the white dwarf in $f(R)$ gravity model against the radius $\mathcal{R}$ are plotted for different values of the parameter $\delta$ and $\xi$. In a single plot $\delta$ is fixed and $M-\mathcal{R}$ curves are obtained for various values of $\xi$.}
\label{fig5}
\end{figure}
\end{enumerate}
\begin{figure}[h]
\begin{center}
\includegraphics[width=0.49\textwidth]{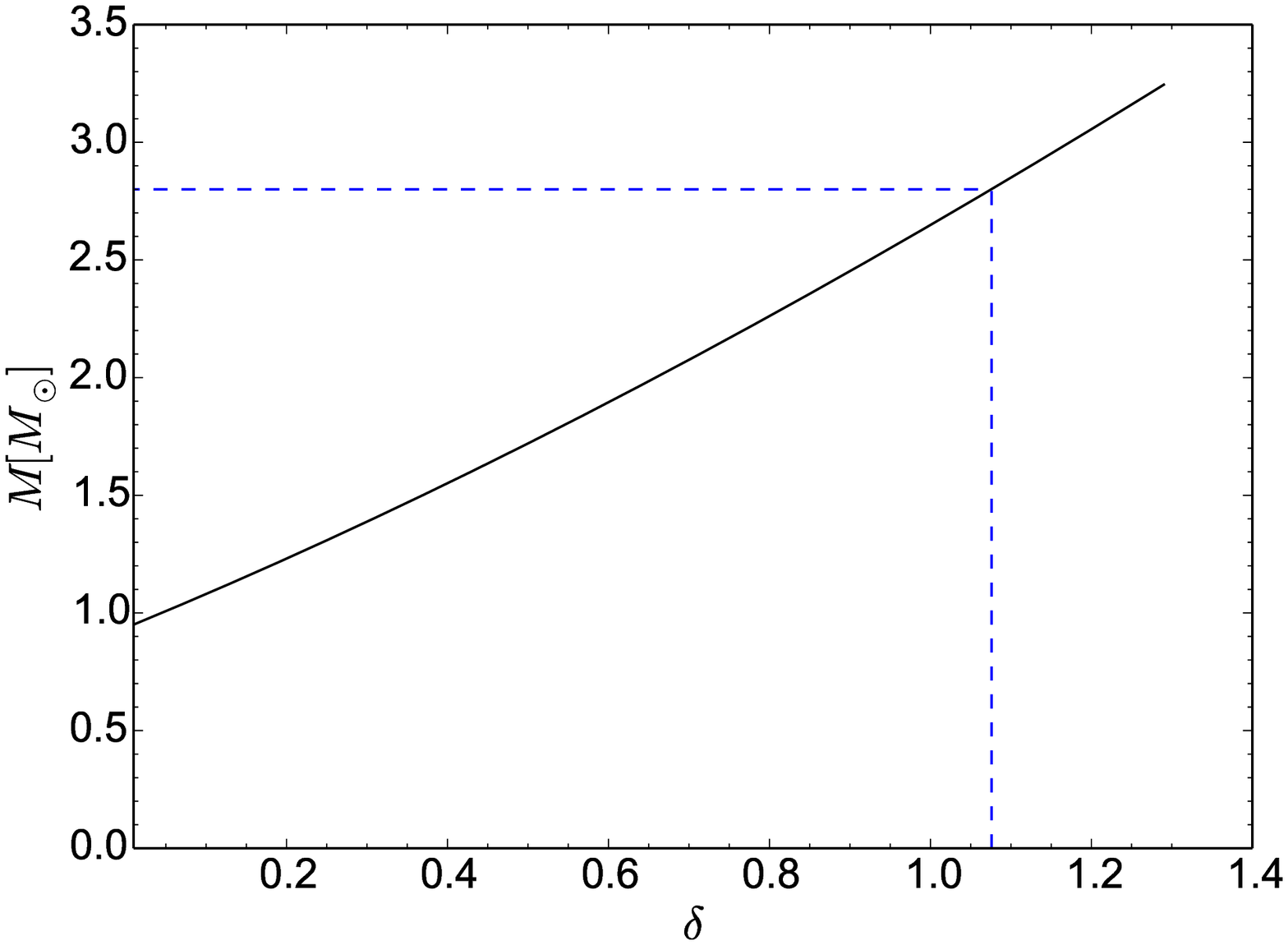}
\caption{Maximum mass $M$ of the white dwarf in $f(R)$ gravity model against the parameter $\delta$. The dotted line corresponds to $M$=$2.8 M_{\bigodot}$.}
\label{fig6}
\end{center}
\end{figure}

All these effects are captured in FIG. \ref{fig6}. Now since gravity is strongest for $\xi=0$ the Chandrasekhar limit would be minimum for any $\delta$. Hence we can effectively constrain the parameter space of $\delta$ by considering the fact that super Chandrasekhar white dwarf mass can have maximum value of $2.8M_{\odot}$ and it produces the constraint $\delta<1.076$.
\subsection{Constraints from observations}\label{Cfo}
In this section, we will constrain the parameters involved in the aforesaid models by comparing their mass-radius relation with a catalog of 12 white dwarfs compiled in \cite{Barstow}, whose masses, radii and the respective errors are known.  This comparison is done using a $\chi^2$ test, with the model parameter(s) as the fitting parameter(s). So we have calculated the $\chi^2$, given by 
\begin{equation}
\Delta \chi_i^2 = \frac{[M-M_i]^2}{\sigma_{M,i}^2} +  \frac{[\mathcal{R}-\mathcal{R}_i]^2}{\sigma_{\mathcal{R},i}^2}
\end{equation}
for each $M$ and $\mathcal{R}$ that we have obtained from our simulation (fixing the parameter values and varying $x_0$) and minimized the quantity for each data point i.e. $(M_i,\mathcal{R}_i)$. Here, $M_i, \sigma_{M,i},\mathcal{R}_i$ and $\sigma_{\mathcal{R},i}$ are mass, standard deviation in mass, radius and standard deviation in radius respectively for the $i$th white dwarf in the catalog. Hence we can obtain the minimum $\chi^2$ by adding the $\Delta \chi^2_i$ for twelve white dwarfs for a fixed set of values of the parameter(s). So one can effectively constrain the parameter space of the modified gravity models at different confidence levels by allowing the reduced $\chi^2$ or $\chi^2_{\nu}$ to take values up to a particular range. This can be best illustrated by plotting $\chi^2_{\nu}$ against the values of the parameter(s) with specifying the region of parameter space up to a particular confidence level. The reduced $\chi^2$ or $\chi^2_{\nu}$ is defined as $\frac{\chi^2}{\nu}$ where $\nu$ is the degrees of freedom. It is given by $\nu=2N-n-1$  where $N=12$ is the number of white dwarfs and the factor of $2$ comes because we have two independent observations, mass and radius. $n$ is the number of fitting parameters. 
\begin{figure}[h!]
\begin{center}
\includegraphics[width=0.49\textwidth]{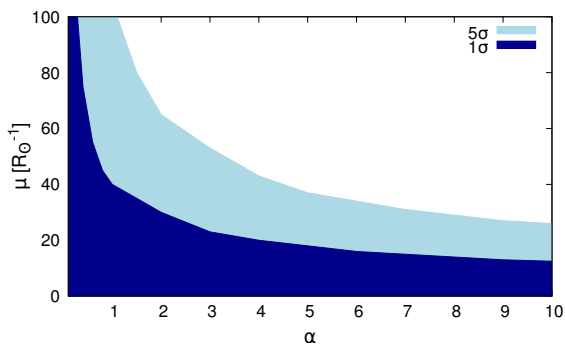}
\caption{The allowed region of parameter space for STVG up to $1\sigma$ (dark blue region) and up to $5\sigma$ (lighter blue region) confidence levels.}
\label{obs1}
\end{center}
\end{figure}

In case of Newtonian gravity i.e. Einstein gravity in Newtonian limit, the $\chi^2_{\nu}$ takes the value $0.73$ and hence one can safely claim that Einstein gravity is in good agreement with the observational data of white dwarf stars. 

\begin{figure}[hb]
\begin{center}
\includegraphics[width=0.49\textwidth]{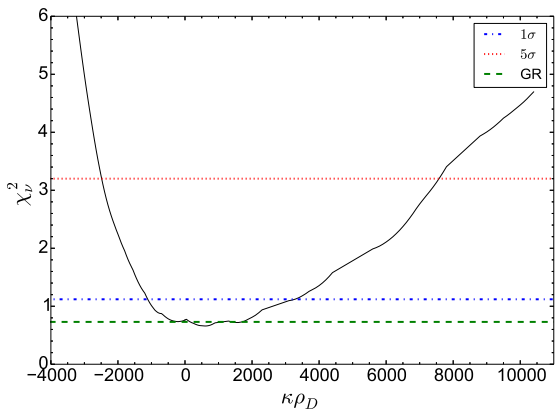}
\caption{$\chi^2_{\nu}$ against the parameter $\kappa\rho_D$ in EiBI gravity. Here $\rho_D = 10^{14} \rho_{\odot}$. The dotted blue line marks the region up to $1\sigma$ confidence level whereas dotted red line marks the region up to $5\sigma$ confidence level. The green dotted line specifies the Newtonian $\chi^2_{\nu}$ value.}
\label{obs2}
\end{center}
\end{figure}
STVG is endowed with two parameters $\alpha$ and $\delta$. Hence d.o.f. is $21$. The allowed region for this model from the observations of white dwarfs is given in FIG. \ref{obs1} where we have mentioned the allowed regions up to $1\sigma$ and $5\sigma$ confidence levels.  Moffat and Rahvar \cite{mof2} have obtained numerical values of the parameters $\mu$ and $\alpha$ by fitting the predicted galaxy rotation curves to observational data. They have got $\alpha=8.89\pm0.34$ and $\mu=0.042\pm0.004$ $\text{kpc}^{-1}$. In our case for $\alpha=8.89$, the allowed region of $\mu$ up to $5\sigma$ confidence level corresponds to $\mu<10 R_{\odot}^{-1}$. Hence the present values of the parameters fall in the allowed region mentioned in FIG. \ref{obs1} which justifies the validity of our analysis. In the limit $\mu \rightarrow \infty$, STVG is strongest where it takes the form of Newtonian gravity with an enhanced factor [See equation (\ref{1f})] and the deviation is only controlled by $\alpha$. In this limit $\alpha <0.183$ is allowed up to $5\sigma$ confidence level. Recently, Armengol and Romero \cite{mof5} considered STVG in the context of neutron stars and put a constraint $\alpha<0.1$ assuming $\mu \rightarrow \infty$. So their predicted parameter space falls within the region we have obtained (i.e. $\alpha <0.183$). But as $\mu$ decreases from $\infty$ the repulsive force increases and hence the upper bound for $\alpha$ also increases (also the white dwarfs would be able to support for mass, see the plots FIG. \ref{fig1}). For example, when $\mu=100 R_{\odot}^{-1}$ the upper bound on $\alpha=1.0$ whereas the upper bound on $\alpha$ is $10$ when $\mu=25R_{\odot}^{-1}$. Hence for any finite value of $\mu$, the region $\alpha<0.183$ is always allowed up to $5\sigma$ confidence level. In the limit $\mu\rightarrow0$ as the repulsive part cancels the enhanced attractive part giving only the Newtonian acceleration, all the values of $\alpha$ are possible quite obviously then. All these features are illustrated in the FIG. \ref{obs1}. 

In case of EiBI gravity, d.o.f is $22$ as there is only one parameter $\kappa$ which can be positive as well as negative. In the limit $\kappa\rightarrow0$, this model coincides with GR. In this case $\chi_{\nu}^2$ is minimized around  $\kappa\rho_D=600$, where $\rho_D = 10^{14} \rho_{\odot}$. Also as $\kappa$ decreases or increases from the above mentioned value the $\chi^2_{\nu}$ value increases (See FIG. \ref{obs2}). One can effectively constrain the parameter space of $\kappa$ from the white dwarf observational data yielding $-1100\leq \kappa\rho_D\leq 2600$ i.e. $-0.7\times10^3m^5 kg^{-1}s^{-2}<\kappa<1.66\times10^3m^5 kg^{-1}s^{-2}$ up to $1\sigma$ and $-2500 \leq \kappa\rho_D \leq 7600 $ i.e. $-1.598\times10^3m^5 kg^{-1}s^{-2}<\kappa<4.858\times10^3m^5 kg^{-1}s^{-2}$ up to $5\sigma$ confidence levels.
\begin{figure}[h]
\begin{center}
\includegraphics[width=0.49\textwidth]{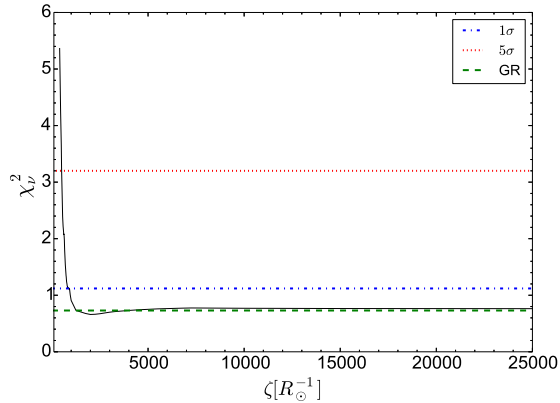}
\caption{$\chi^2_{\nu}$ against the parameter $\zeta$ in FOG model. The dotted blue line marks the region up to $1\sigma$ confidence level whereas dotted red line marks the region up to $5\sigma$ confidence level. The green dotted line specifies the Newtonian $\chi^2_{\nu}$ value.}
\label{obs3}
\end{center}
\end{figure}

FOG model is characterized by the parameter $\zeta$. Hence d.o.f. is $22$. In this case, $\zeta \rightarrow \infty$ is the Newtonian limit. The $\chi^2_{\nu}$ in this model is minimized around $\zeta=2000R_{\odot}^{-1}$ and after that it slowly converges to the Newtonian value as $\zeta$ increases. Also decreasing $\zeta$ from $2000R_{\odot}^{-1}$ increases the value of $\chi^2_{\nu}$. This feature is also illustrated in FIG. \ref{obs3}. Hence one can obtain a lower bound of $\zeta$ from the data. The constraint
we obtain is  $\zeta>900 R_{\odot}^{-1}$ or $L(=\frac{1}{\zeta})<7.73\times10^5 m$ up to $1\sigma$ and $\zeta>500 R_{\odot}^{-1}$ or $L(=\frac{1}{\zeta})<1.39\times10^6 m$ up to $5\sigma$. 
\begin{figure}[h]
\begin{center}
\includegraphics[width=0.49\textwidth]{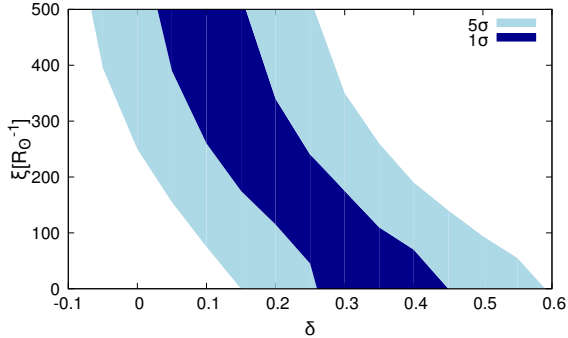}
\caption{The allowed region of parameter space for $f(R)$ model up to $1\sigma$ (dark blue region)and up to $5\sigma$ (lighter blue region) confidence levels.}
\label{obs4}
\end{center}
\end{figure}

As discussed previously $f(R)$ gravity model has got two parameters $\xi$ and $\delta$. Hence d.o.f. is 21. Also the limit $\xi \rightarrow 0$ gives the upper bound on $\delta$ while the limit  $\xi \rightarrow \infty$  gives the lower bound on $\delta$ (See Section \ref{mrf}). Hence we get the constraint $-0.054<\delta<0.450$ up to $1\sigma$ and $-0.155<\delta<0.593$ up to $5\sigma$ confidence levels. In the region described by $\delta\leq0$, as $\xi$ decreases from the limit $\xi\rightarrow\infty$, gravity becomes more and more strong. Hence in this regime there can exist a lower bound on $\xi$ depending upon the value of $\delta$, e.g. when $\delta=-0.1$ the lower bound takes the form $\xi>700R_{\odot}^{-1}$ up to $5\sigma$. Similarly in the region $\frac{1}{3}\leq \delta$, as $\xi$ increases from the limit $\xi\rightarrow0$, gravity becomes more and more weak. Hence there can exist a upper bound on $\xi$ depending upon the value of $\delta$, e.g. when $\delta=0.4$ the upper bound takes the form $\xi<190R_{\odot}^{-1}$ up to $5\sigma$. This behavior is illustrated in Fig. \ref{obs4} showing the allowed region in parameter space. The $f(R)$ gravity model we have chosen would take the form of Starobinsky inflationary model \cite{Star} in the limit $\delta=0$ or $c_1=1$. In this limit we obtain the constraint $\xi>250 R_{\odot}^{-1}$ or $c_2<1.28\times10^{12}m^2$ up to $5\sigma$ confidence level which is nearly of the same order of what one gets from Gravity Probe B experiment giving $c_2<5\times10^{11}m^2$ \cite{Jaf}. Also for $\delta=1/3$ we got the constraint $\xi<350 R_{\odot}^{-1}$ or $\Lambda=\frac{1}{\xi}>1.98\times10^6m$ up to $5\sigma$ confidence level. Recently the case $\delta=1/3$ has been considered in \cite{S2} in the context of orbital precession of S2 star around the supermassive black hole Sagittarius A* and they have found that the most probable value of the parameter $\Lambda$ lies in the range $3000\pm1500 AU$ or $4.5\pm2.25\times10^{14}m$ which falls well within our mentioned parameter space. It can also be shown that the constraint we obtained for $\delta=1/3$ can be successfully applied to the clusters of galaxies \cite{Salzano}.
\begin{table}
\centering
\begin{tabular}{| c | p{5cm} | p{5.8cm}| p{3.1cm} |}
\hline
Model &  \multicolumn{1}{c|}{Modified Poisson's equation} & \multicolumn{1}{c|}{Effective potential} & \multicolumn{1}{c|}{Constraints} \\
\hline\hline
STVG & $\nabla^2\Phi=4\pi G \rho+\kappa\omega\nabla^2\phi_0$ $\nabla^2\phi_0-\mu^2\phi_0=-4\pi\kappa\rho$ & $\Phi\left(\textbf{r}\right)=-G_N\left(1+\alpha\right)\int\frac{\rho\left(\textbf{r}'\right)}{\mid\textbf{r}-\textbf{r}'\mid}d^3r'+G_N\alpha\int\frac{\rho\left(\textbf{r}'\right)}{\mid\textbf{r}-\textbf{r}'\mid}e^{-\mu\mid\textbf{r}-\textbf{r}'\mid}d^3r'$ & For $\alpha=8.89$, $\mu<10 R_{\odot}^{-1}$ \\
\hline
EiBI & $\nabla^2 \Phi=4\pi G_N \rho+\frac{\kappa}{4}\nabla^2\rho$ & $\Phi\left(\textbf{r}\right)=-G_N\int\frac{\rho\left(\textbf{r}'\right)}{\mid\textbf{r}-\textbf{r}'\mid}d^3r'+\frac{\kappa}{4}\rho$ & 
\multirow{1}{*}{$-1.598\times10^3$}
\multirow{1}{*}{$m^5 kg^{-1}s^{-2}$}
\multirow{1}{*}{$<\kappa<0.35$ $\times10^2$}
\multirow{1}{*}{$m^5 kg^{-1}s^{-2}$} \\
\hline
\multirow{1}{*} {FOG} & $\nabla^4\Phi-\zeta^2\nabla^2\Phi=4\pi G_N\zeta^2 \rho$ & $\Phi\left(\textbf{r}\right)=-G_N\int\frac{\rho\left(\textbf{r}'\right)}{\mid\textbf{r}-\textbf{r}'\mid}d^3r'$  &  $L(=\frac{1}{\zeta})<1.45\times$ \\  &  & $+G_N\int\frac{\rho\left(\textbf{r}'\right)}{\mid\textbf{r}-\textbf{r}'\mid}e^{-\zeta\mid\textbf{r}-\textbf{r}'\mid}d^3r'$ & $10^5 m.$ \\
\hline
$f(R)$ & \multirow{1}{*}{$\nabla^2\Phi(\textbf{r})=\frac{4\pi G}{1+\delta}\rho(\textbf{r})-\frac{1}{6\xi^2}\nabla^2R^{(2)}$} \multirow{2}{*}{$\left(\nabla^2-\xi^2\right)R^{(2)}=-\frac{8\pi G_N\xi^2}{1+\delta}\rho$} &
\multirow{1}{*}{$\Phi\left(\textbf{r}\right)=-\frac{G_N}{1+\delta}\int\frac{\rho\left(\textbf{r}'\right)}{\mid\textbf{r}-\textbf{r}'\mid}d^3r'$} \multirow{2}{*}{$-\frac{G_N}{3\left(1+\delta\right)}\int\frac{\rho\left(\textbf{r}'\right)}{\mid\textbf{r}-\textbf{r}'\mid}e^{-\xi\mid\textbf{r}-\textbf{r}'\mid}d^3r'$} & \multirow{1}{*}{$-0.155<\delta<0.593$}
\multirow{1}{*}{For the case $\delta=0$,}
\multirow{1}{*}{$\xi>250 R_{\odot}^{-1}$ or}
\multirow{1}{*}{$c_2<1.28\times10^{12}m^2.$}
\multirow{1}{*}{For the case $\delta=\frac{1}{3}$,}
\multirow{1}{*}{$\Lambda(\frac{1}{\xi})>1.98\times10^{6}m$}\\
\hline
\end{tabular}
\caption{The results for the four modified gravity models are summarized in the above table. For the models EiBI, FOG and $f(R)$ we have got constraints from both the estimation of maximum mass for super Chandrasekhar white dwarfs and observation of white dwarfs mentioned in \cite{Barstow}. But in the table we have only mentioned the best constraint coming from these two observations.}
\end{table}

\section{Conclusions}\label{p}
We explore the effects of four modified gravity theories in the weak field limit by studying the white dwarf stars. We found that all these models leave a rich imprint on the mass radius relation of these stellar objects as a consequence of presence of additional attractive or repulsive terms in the Newtonian limit. Motivated by this, we constrain the parameter space of these modified gravity models from the observation of masses and radii of white dwarfs and we have got reasonable constraint in all the cases.   In all the cases we have observed white dwarfs of either higher or lower masses (compared to Newtonian gravity) depending upon the nature of additional terms (repulsive or attractive). For FOG and EiBI gravity we have found that Chandrasekhar limit does not exist i.e. the mass of white dwarfs does not stabilize to a particular value, rather it increases with increasing central density. We have also considered the super Chandrasekhar white dwarfs in this study and discussed their standing in our chosen modified gravity models. We found that the modified gravity can play an important role in explaining their existence.
\begin{acknowledgments}
We are grateful to Dr. Ameeya A. Bhagwat from CEBS for useful suggestions regarding the numerical scheme we have used in this paper. S.B would like to thank Lankeswar Dey and Krishnendu Mandal from TIFR for helping in preparing the manuscript. 
\end{acknowledgments}
 
\end{document}